\documentclass[sigconf]{acmart}

\usepackage{booktabs} 

\usepackage[T2A,T1]{fontenc}
\usepackage[utf8]{inputenc}
\usepackage[russian,english]{babel}

\usepackage{multirow}
\usepackage{url}
\usepackage{amsmath}
\usepackage{graphicx}
\usepackage{balance}

\def\market{{\texttt {RuMarket}}}
\def\SYM{{\texttt {SYM}}}

\def\NVD{{\texttt {NVD}}}
\def\MAL{{\texttt {MALWARE}}}
\def\STDAL{{\texttt {STANDALONE}}}
\def\AL{{\texttt {STANDALONE}}}
\def\EKIT{{\texttt {EKIT}}}

\begin{document}

\title{Economic Factors of Vulnerability Trade and Exploitation}
\author{Luca Allodi}
\orcid{0000-0003-1600-0868}
\affiliation{%
  \institution{Eindhoven University of Technology}
  \city{P.O. Box 513, Eindhoven} 
  \state{The Netherlands} 
  \postcode{5600 MB}
}
\email{l.allodi@tue.nl}

\copyrightyear{2017}
\acmYear{2017}
\setcopyright{acmcopyright}
\acmConference{CCS '17}{October 30-November 3, 2017}{Dallas, TX, USA}\acmPrice{15.00}\acmDOI{10.1145/3133956.3133960} \acmISBN{978-1-4503-4946-8/17/10}





\begin{abstract}
Cybercrime markets support
the development and diffusion of new attack technologies, vulnerability
exploits, and malware. 
Whereas the revenue streams of cyber attackers have been studied
multiple times in the literature, 
no quantitative account currently exists on the economics
of attack acquisition and deployment. Yet, this understanding is critical
to characterize the production of (traded) exploits, the economy that drives it, and its effects on the overall attack scenario.
 In this paper we provide an empirical investigation of the economics of vulnerability exploitation, and the effects of market factors on likelihood of exploit. Our data is collected first-handedly from a prominent Russian cybercrime market where the trading of the most active attack tools reported by the security industry happens.
 Our findings reveal that exploits in the underground are priced similarly or above vulnerabilities
in legitimate bug-hunting programs, and that the refresh cycle of exploits is slower than currently often assumed. On the other hand, cybercriminals are becoming faster at introducing selected vulnerabilities, and the market is in clear expansion both in terms of players, traded exploits, and exploit pricing. We then evaluate the effects
of these market variables on likelihood of attack realization, and find strong evidence of the correlation between market activity and exploit deployment. We discuss implications on vulnerability metrics, economics, and exploit measurement.
\end{abstract}


\maketitle

\section{Introduction}

The rapid expansion of the cyber-threat scenario reported in the recent literature is fostered by the presence of an `underground' economy that supports the development, deployment, and monetization of cyber-attacks~\cite{hao2015drops,Grier-12-CCS}. 
A few studies analyze the dynamics of the underground and the markets that drive it~\cite{Herley-2010-EISP,hao2015drops,Allodi-TETCS-15}, focusing on either the economic mechanisms that enable the market activity~\cite{Herley-2010-EISP,Allodi-TETCS-15}, or the
`after the fact' analysis of its effects in the real world~\cite{hao2015drops,anderson-2012-WEIS}. However, it remains impossible to fully characterize the underground production of cyber-attacks without a quantitative account of its   economic aspects. For example, 
several allegations currently exist on the costs of an exploit in the underground markets. Similarly, the `economy of malware' is thought
to have significant repercussions on the realization of real-world attacks, yet no scientific account of this relation is currently present in the literature. Likewise, legitimate vulnerability markets~\cite{ruohonen2016trading,zhao2015empirical} have been designed to `compete' with cybercrime markets, but the two systems remain hardly comparable without a more precise account of their economic aspects. In this paper we fill this gap by focusing on the economic aspects of exploit acquisition and deployment, hence providing an additional piece in the otherwise incomplete cybercrime puzzle.

Part of the reason why such analyses are scarce in the literature is the difficulty of identifying and studying `good' or `influential' underground markets. Criminal markets are known to be fraught with economic problems that hinder fair trade, and consequently market development~\cite{Herley-2010-EISP,soska2015measuring}. Moreover, markets trading attack technologies tend to be strongly segregated~\cite{whyforums,soska2015measuring,Allodi-TETCS-15,Sood-IC-2013}, making their access and study more difficult to accomplish. For example, common segregation mechanisms include the implementation of \emph{pull-in} mechanisms, language barriers (especially Russian/Portugese/Chinese), and ingress monitoring~\cite{Allodi-TETCS-15}. 
On the other hand, non-English speaking attackers reportedly generate a significant fraction of attacks~\cite{zhuge2009studying,russian-cybercrime-2011}; this may be partly due to the still
loose international regulation of the cyber-space~\cite{johnson1996law}, as well as economic and social aspects on welfare and higher education in developing countries~\cite{kshetri2010diffusion,dezhina1999science,loyalka2012getting}. For example, Russian cybercriminals are known to produce malware that does not attack ex-Soviet nations (ex-CIS), in an attempt to not
catch the attention of the local authorities. On the other hand, a significant fraction of the malware detected at scale, as well as attack vectors such as exploit kits and booter services~\cite{Grier-12-CCS,Kotov-2013-ESSOS,hutchings2016exploring} are suspected to have been engineered by Russian attackers~\cite{russian-cybercrime-2011,symantec-ekits-2011,Kotov-2013-ESSOS}.

In this paper we characterize the economic aspects
of vulnerability exploits as traded in a prominent Russian cybercrime market (\market\footnote{We do not disclose the real name of the market
to minimize threats to our anonymity.})
for user infections at scale (as opposed to targeted or 0-day attacks), and of their effect on risk of exploit in the wild. Through market infiltration we collect information on trade of vulnerability exploits spanning from 2010 to 2017, and correlate this data with Symantec data on exploits detected at scale. 
Our contribution can be summarized as follows: 

\begin{enumerate}
\item The time between vulnerability disclosure and appearance of exploit in \market\ is shortening, showing that
attackers are becoming more reactive in delivering selected exploits. At the same time, number of actors and of traded exploits is increasing; exploit prices are inflating, and exploit-as-a-service models appear to allow for drastic cuts in exploitation costs. This is particularly relevant for the development of economic models of the underground, and impacts attacker and risk modeling.

\item We find strong evidence of the relation between market activity and likelihood of exploitation in the wild. We find that exploits that spawn higher levels of discussion in \market\ are associated with higher odds of exploitation in the wild, and that high
 market prices hinder exploit deployment at scale. This provides a quantitative link between attacker economics and attack realization, and can directly contribute in the development of (more) realistic attacker models.

\item Exploit prices in the underground markets are aligned with or above those of analogous `legitimate' markets for vulnerabilities and vary between 150 and 8000 USD, whereas the arrival of new exploits is significantly slower than otherwise often assumed. This provides insights on the incentives to participate in the underground economy, and on the dynamics of exploit introduction.
\end{enumerate}

\paragraph{Scope of work} The goal of this paper is to provide insights on exploit economics by analyzing one prominent Russian cybercrime market; this work \emph{does not} aim at providing a
full enumeration of \emph{all} exploits traded in the underground: other cybercrime markets may feature different
sets of exploits and/or foster different cybercriminal activities, as well as enforcing different market regulation mechanisms. Similarly, \market\ does \emph{not} focus on \texttt{0day} exploits, whose employment for attacks at scale is reportedly very limited~\cite{Bilge-12-CCS}, and are outside of the scope of this work.

This paper proceeds as follows. The following section discusses related work. Methodology, data collection, and analysis procedure are discussed in Section~\ref{sec:methodology}. Section \ref{sec:analysis} reports our analysis, and Section~\ref{sec:discussion} discusses our results. 

\section{Related work}
\label{sec:relwork}

The economics and development of underground markets have perhaps been first tackled by Franklin et al.~\cite{Franklin-2007-CCS}. On the other hand, 
Herley et al.~\cite{Herley-2010-EISP} showed that cybercrime economics are distinctively problematic in that the lack of effective rule enforcement mechanisms may hinder fair trading, and as a consequence the existence of the market itself. A few studies analyzed the evolution of cybercrime markets~\cite{whyforums,Allodi-TETCS-15,Grier-12-CCS,Sood-IC-2013,cardenas2009economic,Savage-2011-ICM}, and provided estimates of malware development~\cite{calleja2016look} and attack likelihood~\cite{allodi-RA}, but no quantitative account of economic factors
such as exploit pricing and adoption are currently reported in the literature~\cite{ruohonen2016trading,calleja2016look}. In this paper we provide the first empirical quantification of these economic aspects by analyzing data collected first-handedly from a prominent cybercrime market. 

Recent work studied the services and monetization schemes of cyber criminals,
e.g. to launder money through acquisition of expensive goods~\cite{hao2015drops},
or renting infected systems~\cite{Grier-12-CCS,hutchings2016exploring}. The provision
of the technological means by which these attacks are perpetrated remain however relatively unexplored~\cite{ruohonen2016trading}, with the exception of a few technical insights from industrial reports~\cite{symantec-ekits-2011,russian-cybercrime-2011}. Similarly, a few studies estimated the economic
effects of cybercrime activities on the real-world economy, for example by analyzing the monetization of stolen credit cards and banking information \cite{anderson-2012-WEIS}, the realization of profits from spam campaigns~\cite{Kanich-2008-CCS}, the registration of fake online accounts~\cite{thomas2013trafficking}, and the provision of booter services for distributed denial of service attacks~\cite{karami2013understanding}. However, a characterization of the costs of the technology (as opposed to the earnings it generates), and the relation of trade factors on the realization of an attack is still missing. This work provides a first insight on the value of vulnerability exploits in the underground markets, and the effects 
of market factors on presence of attacks in the wild.

The presence of a cybcercrime economy that absorbs vulnerabilities and generates
attacks motivated the security community to study the devision of `legitimate' vulnerability markets that attract security researchers away from the illegal
marketplaces~\cite{zhao2015empirical}. Whereas several
market mechanisms have been proposed~\cite{Ozment-2004-WEIS,kannan2005market}, their
effectiveness in deterring attacks is not clear~\cite{ransbotham2012markets,Miller-2007-WEIS,kannan2005market}. The so-called responsible vulnerability disclosure is
incentivized by the presence of multiple bug-hunting programs by several providers such as Google, Facebook, and {Microsoft}, or `umbrella' organizations that
coordinate vulnerability reporting and disclosure~\cite{Finifter-2013-Usenix,ruohonen2016trading,zhao2015empirical}. It is however unclear how do these compare
against the cybercrime economy, as several key parameters such as exploit pricing in the underground are currently unknown. Further, it remains uncertain whether the adoption of  vulnerability disclosure mechanisms has a clear effect on risk of attack in the wild~\cite{ransbotham2012markets}. This study fills this gap by providing an empirical analysis of exploit pricing in the underground, and evaluating the effect
of cybercrime market factors on the actual realization of attacks in the wild.

\section{Methodology}
\label{sec:methodology}

Sections~\ref{sec:marketsel} to \ref{sec:collection} present our methodology and provide a detailed description of our data. In Sec.~\ref{sec:analysisprocedure} we outline
the analysis procedure, assumptions, and data handling. Sec.~\ref{sec:bias} discusses observational biases of this study, and Sec.~\ref{sec:ethics} addresses ethical aspects.

\subsection{Market infiltration and evaluation}
\label{sec:marketsel}


\market\ is a forum-based market that can be reached from
the open Internet. Access to the market requires explicit admission by the
market administrators, who validate the access request by performing
background checks on the requester. The main criterion for admission is 
the ability to demonstrate that the requester has control over other identities in affiliated Russian hacking forums, and that he/she has been active in the community.\footnote{Admission criteria were initially enforced upon the 2013 arrest, performed by the Russian authorities, of a prominent market member. 
The enforcement of the admission criteria, albeit still present at the time of writing, has now loosened up.}
Gaining access to \market\ required approximately six months to build a credible 
profile, identify
affiliated markets, and letting our alter-ego gain reputation within the hacking community. 
These activities required some level of proficiency in Russian. Section~\ref{sec:ethics} provides a discussion on ethical aspects. 
As members of the market, we have
access to all the (history of) product information, trades, and prices available to active
participants. 
This analysis spans seven years of market activity (July 2010 - April 2017).

\emph{Criteria for market evaluation.}
It is important to first evaluate whether the selected market is a credible
candidate for analysis, or is yet another example of many `scam-for-scammers' underground forums~\cite{Herley-2010-EISP,Franklin-2007-CCS}. Following indications in the literature on the poor implementation
of cybercrime markets~\cite{Herley-2010-EISP,whyforums}, in previous work~\cite{Allodi-TETCS-15} we performed an analysis of the markets'
economic mechanisms (e.g. addressing information asymmetry~\cite{Akerlof1970}, adverse selection, and moral hazard~\cite{Herley-2010-EISP,eisenhardt1989agency}), traded goods, and participation. We here provide a summary of the considered criteria. A complete account of these aspects is given in~\cite{Allodi-TETCS-15}.

 \emph{Cr.1} \emph{Enforcement of market regulation mechanisms}; market mechanisms enforcing market rules, such as punishment for rippers or presence of trade guarantors or escrows are known to be central to address foundational problems that cripple the economics
 of cybercrime markets~\cite{Herley-2010-EISP,eisenhardt1989agency,Akerlof1970,whyforums,Herley-2011-WEIS} and hinder product quality. We found that in \market, rippers are systematically punished, most sellers use the market escrow services to guarantee transactions, and that the high costs of market entry hinder unfair behavior. 
 
 \emph{Cr.2} \emph{Evidence of trade}. We
 evaluated face evidence of actual trading activity in the market.   
 Accounting for indications from economic literature~\cite{resnick2002trust,Shapiro-BJE-1982}, we investigate trade-related feedback from market participants, discussions in the market threads, product evolution, and type of market interactions; all evidence points toward effective trade mechanisms that foster trading activity.
 
 \emph{Cr.3} \emph{Presence of prominent attack tools reported by the industry}. The relevance of \market\ in the threat scenario is supported by the presence of traders for the most prominent attacks reported by the industry. Among those, we find several exploit kits~\cite{Grier-12-CCS} (e.g. Blackhole, RIG, Eleonore~\cite{symantec-ekits-2011,Contagio}) and malware that led
 numerous infection campaigns (e.g. Zeus, Citadel~\cite{binsalleeh2010analysis,Baltazar-2011-trendlabs,rahimian2014reverse}).


\subsection{Sampling exploits in the underground}
\label{sec:selection}

The unstructured nature of forum-based markets calls for a few additional considerations on data sampling: whereas most (criminal) goods such as drugs, weapons, and illegal pornography can be easily identified and described or demoed by vendors (and therefore measured
by investigators~\cite{soska2015measuring}), the disclosure of too much information on an exploit would destroy its value~\cite{Miller-2007-WEIS}, whereas revealing too little eventually leads to market death (as buyers cannot assess what they buy)~\cite{Herley-2010-EISP}. 
In order to meaningfully sample data points, it is therefore critical to identify the exploit reporting criteria adopted by vendors.

To this aim, we randomly sample 50 posts from \market\ generally referring to \foreignlanguage{russian}{``эксплоит''} (\emph{`exploit'}) and (slang) variants thereof, and evaluate the type of reporting and received market response (i.e. number of replies, and trade evidence~\cite{soska2015measuring}). In our sample we find 19 ads selling 35 exploits overall. Four reporting mechanisms emerge: using the standard Common Vulnerabilities and Exposures (\texttt{CVE}) identifier~\cite{NVD}; describing an exploit as affecting a disclosed vulnerability (\texttt{Knwn}); describing it as a \texttt{0day};  not describing it at all (\texttt{Und.}). 
Table~\ref{tab:idexpl} summarizes the results.
\begin{table}[t]
\small
\centering
\caption{Exploit identification and market activity}
\label{tab:idexpl}
\begin{tabular}{l r r r| r r| r r r}
\toprule
&&\multicolumn{2}{c}{Vulns}&\multicolumn{2}{c}{Replies}&\multicolumn{3}{c}{Trade evidence}\\\cmidrule{3-9}
Type & Ads & Tot. & Avg. & Tot. & Avg.& Yes & No &\% \\\midrule
\texttt{CVE} & 9  & 30 & 3.3 & 518&   57.5 & 5 & 4 & 55\%\\
\texttt{Knwn} & 4 & 4& 1 & 55& 13.8 & 2 & 2& 50\% \\
\texttt{0day} & 1 & 1 & 1 & 44 & 44 & 0 & 1&0\%\\
\texttt{Und.} & 5 & - &-  & 65&13& 1 & 4 &25\% \\
\bottomrule
\end{tabular}
\end{table}
Overall, we find nine adverts reporting 30 CVEs, one reporting a single \texttt{0day}, and four reporting one \texttt{Knwn} vulnerability each. Five additional posts (\texttt{Und.}) advert an undefined number of vulnerabilities without further details on affected software or type
of exploit.
The first observation is that posts reporting CVEs trade on average significantly more vulnerabilities than other posts ($p=0.03$ for a Wilcoxon rank sum test), indicating that this category likely represents the great majority of marketed exploits. 
Adverts reporting CVEs also show greater market activity, 
measured in terms
of received replies ($p=0.05$), than other posts. Similarly, poorly described vulnerabilities are unlikely to show any evidence of trade, whereas \texttt{CVE} and \texttt{Knwn} vulnerabilities display similar rates. Overall, 
we find that vulnerabilities reported by CVE represent the significant 
majority of traded vulnerabilities (by almost an order of magnitude), and receive significantly more attention from the \market\ community than the aggregate.

Further supporting the relevance of CVE reporting, we find that market participants actively look for CVE information when not immediately available; for example, an interested buyer of the KTR package asks (translated from Russian):
``\emph{Which exploits are bundled in the pack at the moment? If possible, specify the CVE}'';
the seller complies. Numerous other examples go in this same direction. 
Critically, this mechanism allows buyers to perform a first assessment
of the exploit, and to verify that the characteristics of the vulnerability it exploits match the vendor's claims (e.g. allow for remote code execution or privilege escalation); this, alongside other market mechanisms described in previous work~\cite{Allodi-TETCS-15}, directly addresses the problem of adverse selection, foundational to all markets of this type, and first underlined in~\cite{Herley-2010-EISP}.
Indeed, vulnerability identification is part
of the regulation of the market itself: for example, a vendor was blocked by the forum administrators when trying to sell (for 3000 USD) an identifiable Windows PoC (\texttt{CVE-2012-0002}); the admin explains: 
``\emph{This [exploit] is public (if not today, tomorrow). The DOS proof-of-concept is already public. Such sales are prohibited}''.

For these reasons in this study we use CVE-IDs as a sampling mechanism for traded exploits. This 
has also the advantage of allowing us to precisely measure
additional characteristics of the vulnerability, including date of disclosure, technical severity, affected software, and presence in the wild, all of which would be impossible without a rigorous definition of published exploits. Importantly, this also rules out errors caused by double counting vulnerabilities, while accounting for the vast majority
of published exploits (\emph{ref. }Tab~\ref{tab:idexpl}). The remaining bias is discussed in Sec.~\ref{sec:bias}.

\subsection{Data collection}
\label{sec:collection}

In this analysis we employ three datasets. The collected data fields are reported in Table~\ref{tab:datvars}; in the Appendix we report an extended description of each field.
\begin{table*}
\small
\centering
\caption{Summary of collected variables used in the analysis}
\begin{minipage}{0.95\textwidth}
\footnotesize
Description and summary statistics of the collected data fields. \emph{Unit} indicates the type of data field.
\emph{Lvls} indicates, for categorical variables, the number of factor levels.
Descriptive statistics are provided for cardinal/ordinal variables, and 
categorical variables with only two factors (encoded as 1: presence of condition; 0: absence of condition).
From the product descriptions naturally emerge the following package categorization (vulnerability descriptions from \texttt{NVD}~\cite{NVD}):

a) \AL: packages traded as stand-alone exploits that are then personalized by the buyer. For example, \texttt{CVE-2016-0189} ``\emph{allow[s] remote attackers to execute arbitrary code [..] via a crafted web site}'', and is traded in \market\ as a \AL\ exploit to which the customer can add his/her own `\emph{private}'  (sic.) shellcode.

b) \MAL: exploits embedded in malware packaging services. Exploits in these packages typically allow for privilege escalation. For example, \texttt{CVE-2015-1701} ``\emph{allows local users to gain privileges via a crafted application}'', and in \market\ is bundled in a \MAL\ dropper that, when executed on the target machine, escalates to higher privileges and executes the custom code.

c) \EKIT: exploit packages typically rented (as opposed to traded) as exploit kits, namely  web servers that deliver exploits and custom payloads to victims that are redirected to the kit~\cite{Kotov-2013-ESSOS}. The rental period in our sample goes from a week to a month. \EKIT s operation requires the execution of arbitrary code on the victim system to remotely drop the malware. For example, \texttt{CVE-2016-1019} ``\emph{allows remote attackers to [..] execute arbitrary code}'', and in \market\ it is embedded in the notorious exploit kit \texttt{RIG}.
\end{minipage}
\label{tab:datvars}
\begin{tabular}{ p{0.17\columnwidth} p{0.17\columnwidth} l  p{0.59\columnwidth} r r r r r}
\toprule
Variable & Dataset & Unit  & Description &Lvls& Min & Mean & Max & sd\\
\midrule 
\texttt{CVE} & \market, \NVD, \SYM & Cat. & The unique identifier of the  vulnerability. & 57 &&&&\\
\texttt{CVEPub} & \NVD & Date & Date of vuln disclosure in \NVD. &  &2006-04-11&2012-12-06&2016-11-10&970.84\\

  \texttt{ExplVen}  & \market & Cat.  & The identifier of the product vendor. & 23&  &  &  & 
  \\
 \texttt{ExplVenReg} & \market & Date & Date of vendor registration in the market. & & 2008-05-25 & 2012-10-31 & 2016-03-27 & 873.05
\\
 \texttt{Pack} & \market & Cat. & Bundle of exploits traded in the market. & 38 &  &  &  & 
\\
 \texttt{PubDate} & \market & Date & Date of exploit introduction in a package. &  & 
 2010-07-29 & 2014-06-25 & 2017-01-19 & 640.02
\\
 \texttt{PackType} & \market & Cat.& Pack classification in one of the categories \AL; \EKIT; \MAL. & 3&  &  &  & 
\\
 \texttt{PackPrice}& \market & USD & Acquisition cost of the package. & & 100 & 2417 & 8000 & 2408.28
\\
 \texttt{PackActiv}& \market & Messages & Number of responses
 to package advert. & & 0 & 43.85 & 300 & 69.29
\\
 \texttt{PackDeath}& \market & Date & Date of last reply for the package. & & 2010-12-24 & 2015-03-16 & 2017-03-27 & 642.63
\\
 \texttt{ExplPrice} & \market & USD & Price estimate of single exploit. && 13.64 & 969.00 & 8000 & 1708.76
\\
  \texttt{SwVen} & \NVD & Cat. & Vendor of the vulnerable software.& 3&  & &&\\
 \texttt{Sw} & \NVD & Cat. & Name of the affected software.& 7&  & &&\\
\texttt{CVSS} & \NVD & Ord. & Vulnerability severity measured by the 
Common Vulnerability Scoring System.&& 5  &8.76 &10&1.33\\

 \texttt{SYM} & \SYM & Cat. & Presence of exploit at scale. &2& 0 & 0.84 & 1 &0.38\\
\bottomrule
\end{tabular}
\end{table*}

\begin{enumerate}

\item \market. We query \market\ and analyze results by reading 
discussion topics and extrapolating relevant information. Unfortunately the nature of the data limits the applicability of fully-automated data extraction procedures (e.g. product updates and multiple products per advert, see also~\cite{portnoff2017tools}). We therefore employ semi-automated pattern matching and manual analysis to extract the information.
We identify traded CVEs by querying \market\ for matches to the case-insensitive regular expression \texttt{cve(-id)?(?i)} in the \emph{Virus, attacks, and malware} commercial section of the forum.  
This procedure returned 194 discussion threads and approximately 3000 posts
to examine in April 2017.
 To minimize the chances of reporting `fake' exploit products, we consider only vendors that have not been reported as `rippers' or banned from the community. This leaves us with a sample of 89 traded exploits over 57 
 unique vulnerabilities embedded in 38  packages for \AL, \MAL\ and \texttt{ExploitKit} products, and attacking \texttt{Microsoft}, \texttt{Oracle},\footnote{All vulnerabilities labeled as \texttt{Oracle} are relative to the Java platform. Some of those were disclosed while Java was Sun's.} and \texttt{Adobe} software. This is quantitatively in line with previous studies on marketed exploits~\cite{Contagio,Allodi-2014-TISSEC,Kotov-2013-ESSOS,ablon2014markets}. 

\item \NVD. The National Vulnerability Database~\cite{NVD}, is the NIST-maintained vulnerability
database reporting vulnerability characteristics,
affected software, and severity.

\item \SYM. Vulnerabilities for which Symantec's threat explorer 
and attack signature databases report an exploit in the wild~\cite{Dumitras-11-BADGERS}. Vulnerabilities outside
of \SYM\ may still be actively exploited, but are unlikely to be
exploited at scale~\cite{Dumitras-11-BADGERS,Allodi-2014-TISSEC}. 
This allows us to correlate 
technical and market characteristics
of vulnerabilities to the actual (mass) realization of an observed exploit in the wild.
\end{enumerate}

We join the three datasets on the CVE-ID of the vulnerability.
%



\subsection{Analysis procedure}
\label{sec:analysisprocedure}


\subsubsection{Estimation of exploit prices}

When a
package contains more than one exploit, the cost of a single exploit can only
be estimated. From the literature
on exploit development and deployment~\cite{erickson-2008-hacking,demott2015bypassing,Zhao-SIW-2014,Finifter-2013-Usenix,Allodi-2014-TISSEC,Kotov-2013-ESSOS} two aspects of vulnerabilities emerge
as drivers of exploitation effort: 1) vulnerability type   
(e.g. memory corruption vs cross-site-scripting)~\cite{erickson-2008-hacking,Kotov-2013-ESSOS,Zhao-SIW-2014}; 2) exploitation complexity (e.g. to evade attack
mitigation techniques)~\cite{demott2015bypassing,erickson-2008-hacking,BOZORGI-etal-10-SIGKDD}.

\emph{Vulnerability type.}
The MITRE corporation maintains a community-developed standard (Common Weakness Enumeration, CWE in short) for the enumeration of
software weaknesses~\cite{NVD}. 
This has the purpose of identifying the type of technical issue that generates
the vulnerability.
Table~\ref{tab:cwebypackage} in the Appendix provides a detailed breakdown of vulnerability CWE types by exploit package in \market. We find that packages typically embed vulnerabilities of the same type (e.g. either remote code execution vulnerabilities or privilege escalation vulnerabilities), which 
suggests that significantly
skewed distributions of exploitation efforts by vulnerability type within a package are unlikely.

\emph{Exploit complexity}. The Common Vulnerability Scoring System (CVSS)~\cite{first-2015-cvss3} defines \emph{Access Complexity} as a measure of whether a reliable exploit
can be `easily' obtained or additional measures or attacks are required to, for example, avoid
attack mitigation techniques (memory randomization, canaries, etc.), 
 or address specific software/system architectures~\cite{first-2015-cvss3,erickson-2008-hacking}. 
 CVSSv2 assesses attack complexity in three categories: High, Medium, Low~\cite{SCAR-MELL-09-ESEM}. 
\texttt{AC:High} conditions have been shown to represent a threshold
for exploit adoption~\cite{Allodi-2014-TISSEC}, whereas Medium and Low vulnerabilities
require only limited exploitation efforts~\cite{SCAR-MELL-09-ESEM} and are commonly
detected in the wild~\cite{Allodi-2014-TISSEC,BOZORGI-etal-10-SIGKDD}. 
Acknowledging this, the newer version of CVSS (v3) considers only High (existence of conditions outside
of the attacker's control) or Low (absence of conditions) values~\cite{first-2015-cvss3}.
Out of 57 unique CVEs in our sample, we find 2 vulnerabilities characterized by a High CVSS attack complexity, whereas the remaining 55 include only
limited or no exploitation complexities for the attacker. A further breakdown
of attack complexity by package (Table~\ref{tab:cvsspack} in the Appendix) shows that
most packages prevalently 
include vulnerabilities with the same \texttt{AC} assessment. This once again suggests that 
exploit development efforts are not significantly skewed among vulnerabilities bundled in a package.

In light of these considerations, in this study 
we estimate unitary cost of exploit by assuming a uniform distribution of costs among exploits in a package.\footnote{Sec.~\ref{sec:exploits} gives a detailed account of how this relates to different software packages.}

\subsubsection{Bootstrapped analysis of exploit prices} In an effort to provide a more precise estimate of exploit costs, 
we employ a block bootstrap analysis ($N=10000$) of exploit prices. 
 The bootstrap procedure randomly re-samples ($N$ times), with replacement, exploit
 packages (i.e. our `blocks', or sampling units) from \market, and approximates
 the true unknown distribution of the  population of traded exploits (of which we observe a sample)~\cite{Efron-1994-bootstrap}. This allows us to infer the parameters of the true
 distribution and to build
 robust confidence intervals of price estimates.

\subsubsection{Regression analysis} The nature of the sample requires a few additional precautions to be taken for a formal analysis. 
In particular, our exploit observations depend not only on the exploit, but also on the specific
vendor who publishes the package where the exploit is bundled in. For example, qualified
vendors may publish more reliable exploits that are more likely to generate attacks in the wild. Hence, the measure of an exploit implicitly depends on the vendor who publishes it (i.e. our data has an hierarchical structure~\cite{agresti2011categorical}).\footnote{\texttt{ExplVen} and \texttt{Package} are both meaningful levels in the hierarchy
of our sample. As it is the vendor of the exploit that publishes the
exploits, fixes exploit cost, and determines exploit quality, we here consider the vendor
as the main source of variance.} This `mixed effect' should be captured to assure
an unbiased quantitative analysis.
We denote $\mu_a$ as the (univariate) random effect for the vendor $a$
such that the expected value of the observation for the $i^{th}$ measurement is
$E(Y_{ai}|\mu_a)=\mu_{ai}$, i.e. the expected value for the $i^{th}$ observation
is conditional on $\mu_a$. The general regression form of our analysis is derived from~\cite{agresti2011categorical} and is:

\begin{equation}
f(\mu_{ai}) = z_{ai}\mu_a + \boldsymbol\beta \boldsymbol x_{ai}
\end{equation}

where $f$ is the link function, $z_{ai}\mu_a$ quantifies the random
effect at the intercept,
 and $\boldsymbol\beta \boldsymbol x_{ai}$  is the vector of
fixed effects and respective coefficients. Standard model diagnostics are run for all regressions.
We report model Log-likelihoods for model comparison. The calculation of $p$-values and model power
are not straightforward for mixed effects models. We report $pseudo-R^2$ and $p-values$ 
as approximations provided by the \texttt{R} packages
\texttt{lmerTest} and \texttt{MuMIn}, alongside the standard deviation of coefficient estimation.

\subsection{Limitations}
\label{sec:bias}

The adopted \texttt{CVE} sampling mechanism may exclude some potentially relevant vulnerability that we cannot measure precisely. Results in Tab~\ref{tab:idexpl} indicate that this sampling bias is likely minimal. It is however worth noting that this is, unfortunately, an inherent limitation of all studies on this type of markets: without engaging in the trading activity, it is impossible to reliably measure what lies behind a market post. For example the excellent work in~\cite{soska2015measuring} conservatively estimates market size by assuming that user feedback relates to separate, single trade lots, as it is not possible to measure multiple trades in a single transaction. Similarly, as we cannot measure unidentified exploits, our analysis should be considered a conservative estimate of traded exploits in \market.

The data collection in \SYM\ reports exploits deployed \emph{en-mass} against consumer (typically Windows) systems, and does not directly extend to targeted attacks and 0-day vulnerabilities.

\subsection{Ethical aspects and data sharing}
\label{sec:ethics}

The market infiltration was performed while the author was as the University of Trento, Italy.
All data collection happened at the Eindhoven University of Technology, the Netherlands. No activity involved the deception of market
participants other than for our `identity'. We only engaged in discussion on non-hacking topics not to facilitate illegal activities. We use the anonymous network TOR to conceal our identities. To preserve our anonymity in the market we do not disclose the real name of the community.
The collected data is available for sharing.\footnote{Access procedure  available at \url{http://security1.win.tue.nl}.}

\section{Data analysis}
\label{sec:analysis}
This analysis is structured in three parts. In the first (Sec.~\ref{sec:overview})
we describe the market  by analyzing the activity of market participants and the characteristics of the traded exploit packages. In the second part (Sec.~\ref{sec:exploits}) we analyze market factors driving exploit prices, and in the third (Sec.~\ref{sec:sym}) the adoption of exploits in the wild. 

\subsection{Overview of \market}
\label{sec:overview}

\subsubsection{Exploit vendors}
\label{sec:vendors}

\begin{figure}
\centering
\includegraphics[width=0.8\columnwidth]{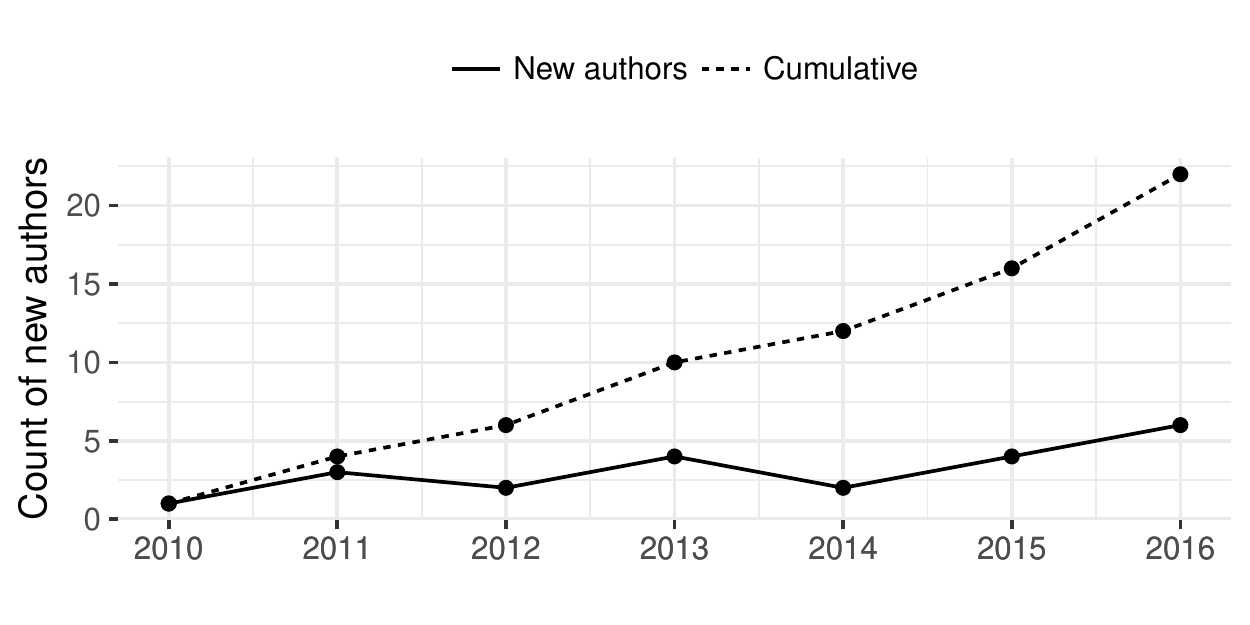}
\caption{Count of vendors trading CVEs in \market}
\label{fig:authappearance}
\end{figure}

\begin{figure}
\centering
\includegraphics[width=0.45\columnwidth]{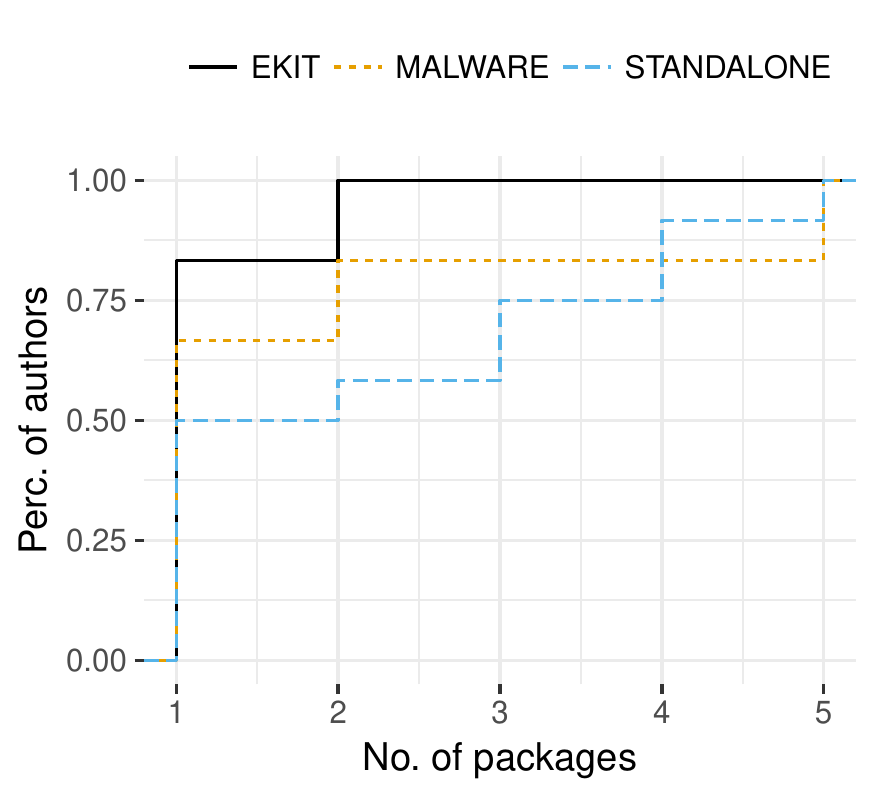}
\includegraphics[width=0.45\columnwidth]{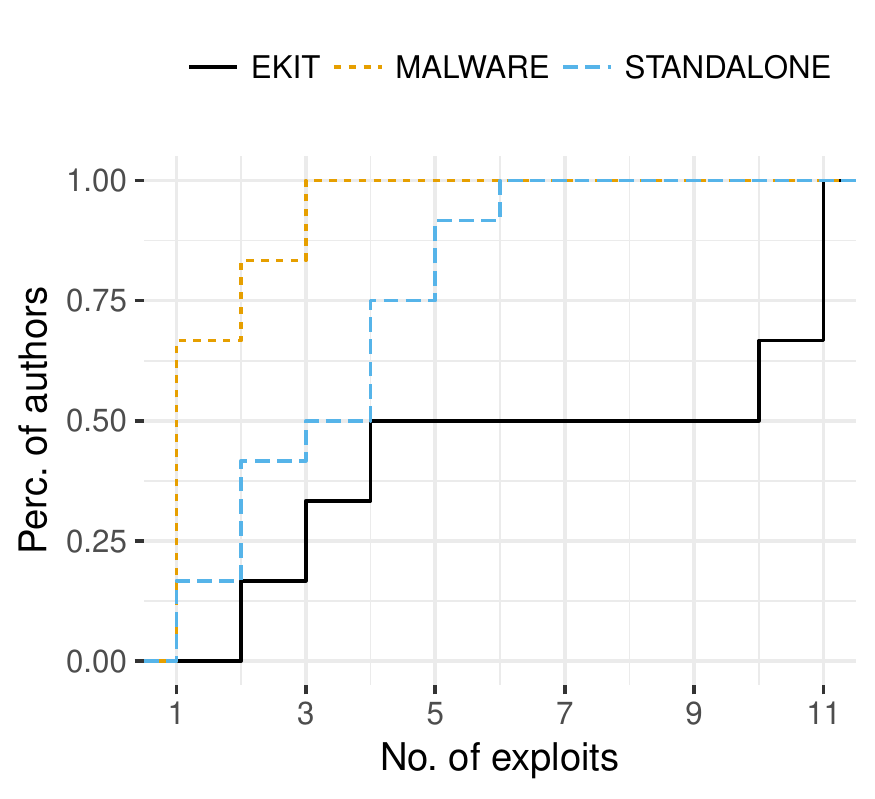}
\caption{Distribution of packages and exploits by vendor}
\label{fig:vendorexploitsvspacks}
\end{figure}

It is first important to provide an overview of the exploit vendors that participate in \market\ activities. Our \market\ sample contains 22 uniquely identified vendors that 
trade CVEs in \MAL, \AL, and \EKIT\ packages. The market mechanism generates strong
disincentives for the creation of multiple accounts~\cite{Allodi-TETCS-15}. Following the approach adopted for similar applications in related work~\cite{soska2015measuring}, in the following we consider vendor aliases as unique seller identifiers. Figure~\ref{fig:authappearance}
shows the appearance of vendors trading exploits in \market. The solid line
reports the count of  new vendors appearing in the market 
(i.e. vendors that did not publish
a CVE exploit in the preceding years in \market\ under the same alias); the dotted line
reports the cumulative count of vendors. The number of vendors increases at a steady linear
rate of approximately three new vendors per year during the observation period. This suggests that exploit trading in \market\ is growing.
Figure~\ref{fig:vendorexploitsvspacks}
offers a breakdown of vendor activity by product type by plotting the CDF of number of exploit
packages introduced by each vendor and the number of exploits they embed in their products.
\EKIT\ vendors typically publish only one product, whereas \MAL\ and particularly \AL\ vendors
appear to trade significantly more packages. This is interesting to observe as \EKIT\ products (and to a lesser degree \MAL\ products~\cite{rahimian2014reverse}) are traded under the `exploit-as-a-service'
model, whereby the seller maintains a service for a period of time during which customers
rent the kit to deliver their own attacks. The maintenance operations include delivering vulnerable traffic to the customers, updating the exploit portfolio, and packing existing exploits to minimize detection in the wild (to generate the so-called \emph{FUD}, `\emph{Fully UnDetectable}' exploits)~\cite{Kotov-2013-ESSOS}.
The implied prolonged contractual form explains the prevalence of vendors
with only one exploit package in their portfolio for \EKIT\ and \MAL\ vendors. On the other hand, \EKIT\
vendors are by far the more `productive' in terms of number of exploited vulnerabilities, with 50\%
of \EKIT\ vendors contributing more than 10 exploits. \AL\ vendors typically focus on a few exploits only, trading on average below three exploits, and only a small fraction of vendors trades overall more than 5 exploits. \MAL\ vendors are the least productive in terms of exploited CVEs: whereas historically Internet worms and malware such as Slammer or Conficker exploited software vulnerabilities to replicate, 
in recent years infections
happen mostly through Malware Distribution Networks~\cite{Goncharov-2011-traff,Grier-12-CCS,Provos-2008-USENIX} that implement the target exploitation by other means (e.g. exploit kits or `malvertising'), and allow for the malware to be `dropped' on the attacked system, with only a few exceptions.

\begin{figure}
\centering
\includegraphics[width=0.45\columnwidth]{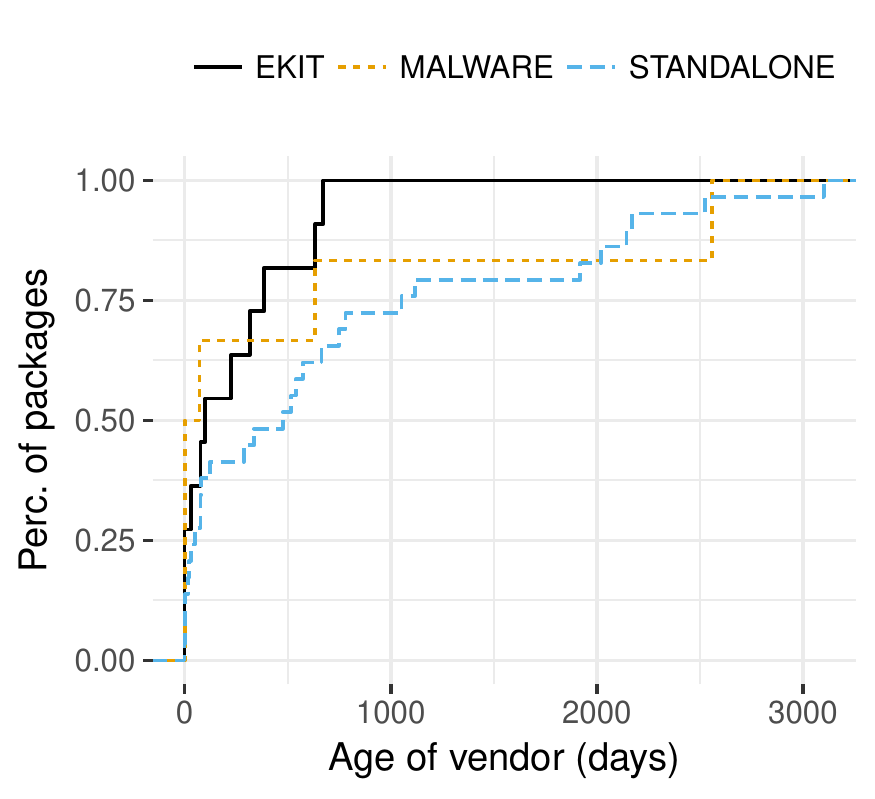}
\caption{Distribution of exploit vendor age}
\label{fig:vendorDOn}
\end{figure}

Foundational
studies in economics~\cite{Shapiro-BJE-1982} as well as more recent research on 
online marketplaces~\cite{Cabral-JIE-2010} put the emphasis on the relation between (expectation of) product quality and placement of the vendor in the market. Due to the unreliability of user feedback on online forums,
criminal online markets
often employ as a proxy for trustworthiness criteria such as time-on-market or 
number of messages/specialty~\cite{holt2016examining,Allodi-TETCS-15}. These are costlier for malicious vendors to replicate than simply posting positive feedback
on their own products.
We use as a proxy measure of seller presence in the market the number of days the vendor have been registered on \market\ at the time of package publication, and calculate it as $\texttt{ExplVenAge}=\texttt{PubDate}-\texttt{ExplVenReg}$.
Figure~\ref{fig:vendorDOn}
reports this distribution by package type. Exploit vendor age varies considerably by type of package. \STDAL\
vendors are those with the highest average time on market at time of product publication.  50\% of \STDAL\ vendors
have been registered on the market for at least six months, whereas only the top 30\% of \EKIT\ and \MAL\
vendors are above this threshold. Overall, we find that only 18\% of vendors publish their first package on the day of registration. 55\% of vendors have been registered for at least a month, and 32\% for at least a year. This indicates that \market\ mechanisms encourage prolonged market activity, which
may determine higher levels of trust among market participants~\cite{holt2016examining}.

\subsubsection{Exploit packages}
\label{sec:packages}

\begin{figure}
\centering
\includegraphics[width=0.8\columnwidth]{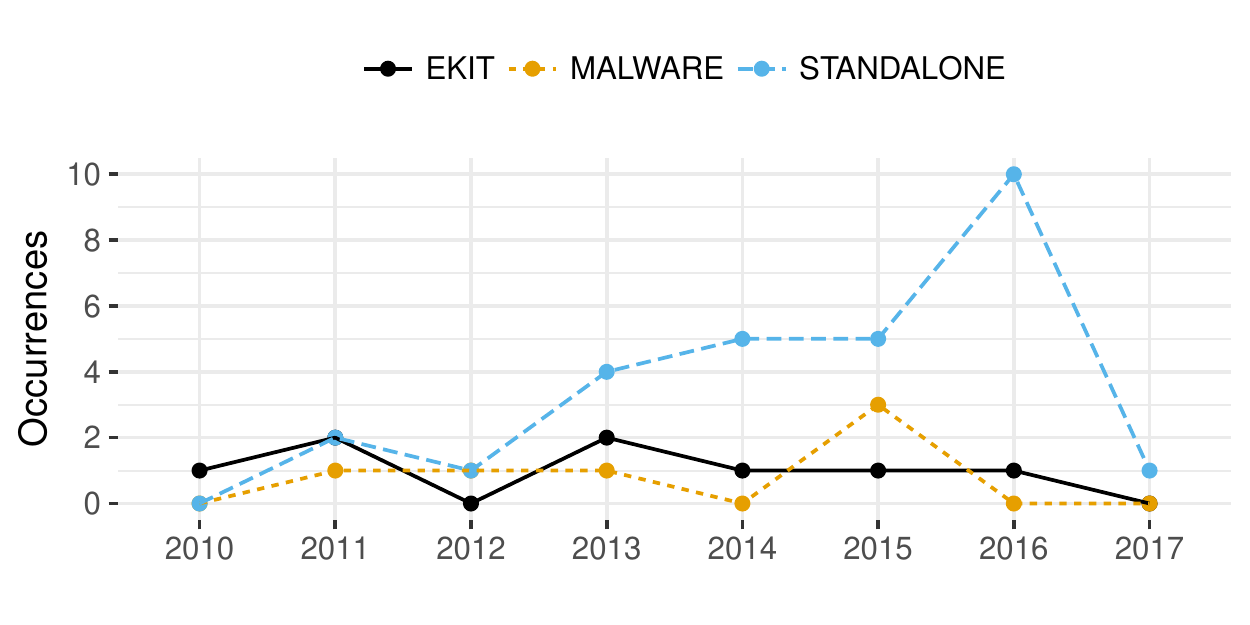}
\caption{Release of exploit packages by type per year}
\label{fig:typeYear}
\end{figure}

\begin{figure}
\centering
\includegraphics[width=0.45\columnwidth]{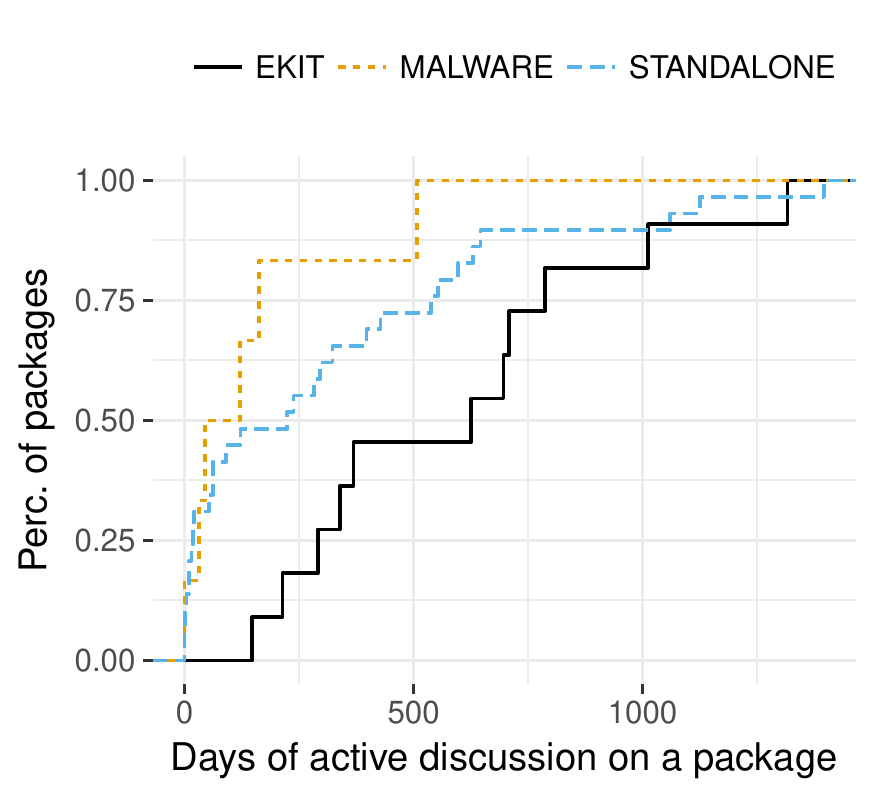}
\includegraphics[width=0.45\columnwidth]{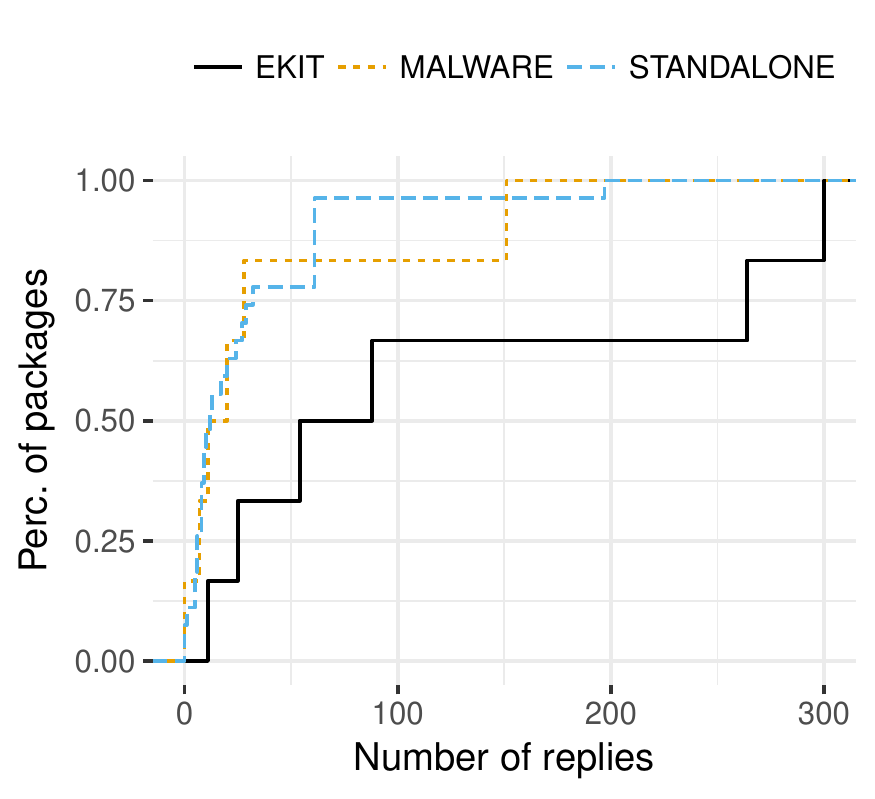}
\caption{Market activity by exploit package}
\label{fig:aliveType}
\end{figure}

Our \market\ sample reports data on 38 unique exploit packages; the breakdown is
as follows: six \EKIT, six \MAL,
and twenty-six \AL\ packages. We consider the addition of new exploits
in a pack as an update to an existing package.
Figure~\ref{fig:typeYear}
reports the number of updates to the exploit package portfolio by year in the market. In general, we can observe that the number of released products steadily increases every year. This trend appears to be mainly driven by \STDAL\ packs, whereas
\EKIT\ and \MAL\ packages are essentially stable in time, with the latter being the lowest on average. This is in line with the exploit authors' activity
described above, and suggests that these packages may enjoy a longer
activity in the market. 
To evaluate this, we consider the days and volume of active discussion since publication 
in the market as a proxy measure of \market's  interest in the product. We compute days of active discussion as $\texttt{\texttt{DaysActive}=\texttt{PackDeath}-\texttt{PubDate}}$, and report $\texttt{PackActivity}$ for volume of active discussion.
Figure~\ref{fig:aliveType} plots the two distributions.
The left plot reports longevity of \market\ activity around an exploit package. Inspection of the market message board reveals that the length
of activity around a package is not artificially inflated by the package vendor 
by continuously adding comments to the advert. As expected, we observe that 
longevity of discussion around \EKIT\ packages is significantly higher than for \STDAL\ 
and \MAL\ packages ($p=0.015$ and $p<0.01$ respectively for a Wilcoxon rank sum test).
\market\ discussion around \EKIT\ packages remains active for more than 500 days (approximately a year and a half) for 50\% of packages, with the top 10\% products
remaining active in the market board for more than 3 years. 
Differently, 50\% of \AL\ and \MAL\ packages remain active for up to approximately 220 days,
and only less than 25\% remain active for more than 500 days. Overall, we find that the average package remains active for a year since time of publication.
The right plot in Figure~\ref{fig:aliveType} plots the distribution of replies by package type. \EKIT\ packages receive on average significantly more replies than other pack types, which is in line with previous figures. Conversely, \market\ interest around \AL\ and \MAL\ packages
is significantly lower, with only a handful of packages receiving a comparable volume
of discussion as the average \EKIT. The lower interest of the \market\ community may be driven by the higher difficulty of use of \AL\ and \MAL\ products, that require additional effort 
to deploy and deliver the attack compared to \EKIT\ products~\cite{Grier-12-CCS}. Further, different price-tags, investigated below, may explain the overall market interest.

\paragraph{Exploit pack prices} Table~\ref{tab:packprices}
\begin{table*}
\small
\centering
\caption{Descriptive statistics of package prices and bundled exploits}
\label{tab:packprices}
\begin{tabular}{l rrrrrrrr|rrrrrrr}
\toprule
& \multicolumn{7}{c}{Package price (USD)} & \multicolumn{7}{c}{no. bundled exploits} &\\
\cmidrule{3-16}
Type&n & Min&0.025p&Mean&Median&0.975p&Max&sd&Min&0.025p&Mean&Median&0.975p&Max&sd\\
\midrule
\EKIT&6&150&157.92&693.89&400&1875&2000&708.94&2&2.12&6.83&7&11&11&4.26\\
\MAL&6&420&428.75&1735&1250&3875&4000&1456.38&1&1&1.5&1&2.88&3&0.84\\
\AL&26&100&100&2972.69&3000&8000&8000&2629.39&1&1&1.5&1&4&4&0.86\\
\texttt{All}&38&100&100&2417.46&1500&8000&8000&2408.28&1&1&2.34&1&11&11&2.63\\
\bottomrule
\end{tabular}
\end{table*}
reports descriptive statistics of exploit pack prices and number of
bundled exploits by package type. 
\AL\ packages are traded at a mean price
around 3000 USD up to 8000 USD, and bundle in between 1 and 4 exploits. The small
standard deviation indicates that most \AL\ packages bundle 1 exploit only.
\MAL\ packages are traded at a price range between 400 USD and 4000 USD, with most
package prices set at around the 1000-2000 USD mark. Similarly to \AL\ packages, \MAL\
bundles typically include only one exploit, and up to three exploits. Finally, the
 lower 50\% of \EKIT s are priced (accounting for an average rent of 2-3 weeks~\cite{huang2014socio}) in the range 150-400 dollars, whereas the upper 50\% are in the range 400-2000 USD. \EKIT\ packages embed significantly more exploits than other package types. This is in line with previous findings in the literature~\cite{Kotov-2013-ESSOS} and, following Grier et al.~\cite{Grier-12-CCS}, this allows for a greater flexibility in terms of the range of selectable targets~\cite{Allodi-ECIS-15}. We give an account of the specific exploits in the next section.

\begin{figure}
\centering
\includegraphics[width=0.8\columnwidth]{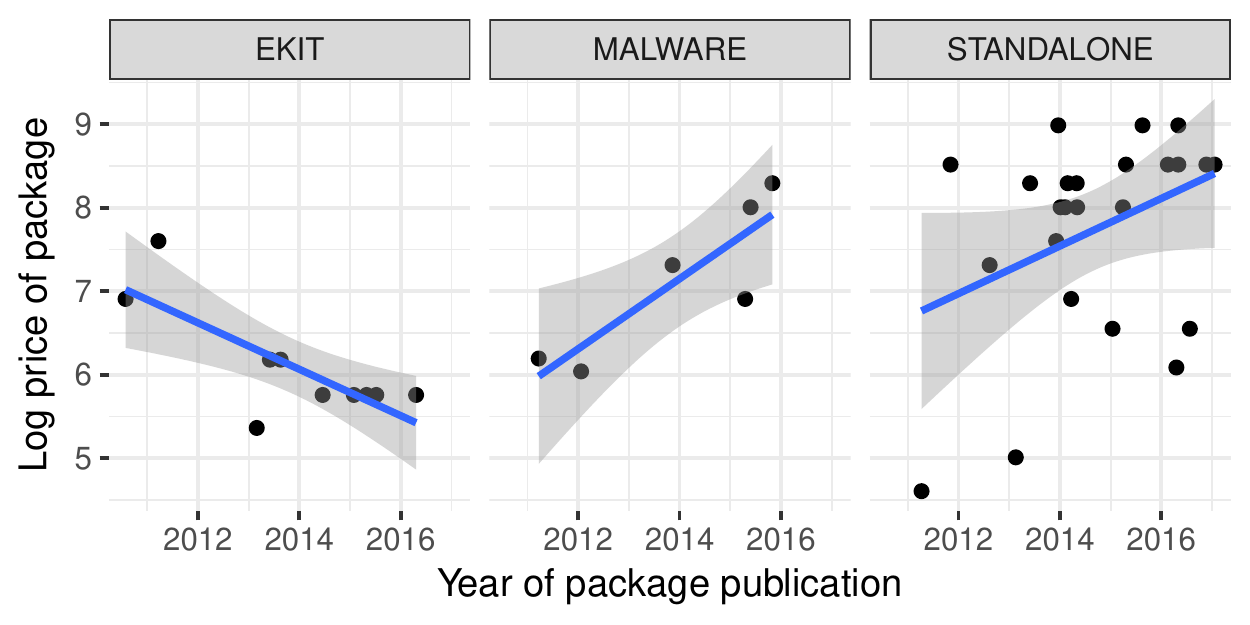}
\caption{Log package price by year by product type}
\label{fig:packpricevsyear}
\end{figure}

Following~\cite{ruohonen2016trading}, we 
further investigate possible outliers in our data to mitigate pricing noise. We find that the only 
four \AL\
packages that received no trade reply from the \market\ community were also traded at below average prices (in between 100/300 dollars each in year 2016). Similarly, we find only one \EKIT\ that, despite embedding twelve exploits, is priced at 150 dollars, significantly below the \EKIT\ mean of 560 USD.
Figure~\ref{fig:packpricevsyear}
plots the results in terms of trend in price per product type by year
with these outliers removed. \MAL\ and \AL\ packages show an increasing trend whereas \EKIT\ product prices are steadily decreasing. 
Regression coefficients for the
 linear model displayed in Figure~\ref{fig:packpricevsyear} are significant at the 5\% level for 
 \MAL\ ($\beta=0.43,\ p=0.05$) and \EKIT\ ($\beta=-0.28, p=0.02$) but not significant for
 \AL\ packages ($\beta=0.23,\ p=0.14$).
 We find that `consumer' services
such as \EKIT\ products are becoming more easily available to the users, a figure compatible
with the increasing trend of `commodified' attacks delivered in the wild~\cite{Grier-12-CCS,hutchings2016exploring,Nayak-2014-RAID,Allodi-ESSOS-15}, whereas the remaining more `specialized' sector of the market seems to be inflating.
We do not find any significant association between number of exploits in the package and package price. This lack of correlation may indicate that the business
model behind exploit trading, as well as other contextual considerations
on market status, presence of similar exploit, and affected software should be considered in the analysis, as previously suggested by several authors~\cite{ruohonen2016trading,Grier-12-CCS,Anderson:2008,eeten-2008-oecd,asghari2013security,Allodi-ECIS-15}. We give an extended account of this in the next section.

\subsection{Analysis of exploits}
\label{sec:exploits}

\subsubsection{Exploit demographics} Embedded in the packages we find 89 exploits targeting 57 unique CVEs in \texttt{Microsoft}, \texttt{Adobe}, and \texttt{Oracle} products. 
Table~\ref{tab:vendorswtype}
\begin{table}[t]
\small
\centering
\caption{Breakdown of traded vulnerabilities}
\label{tab:vendorswtype}
\begin{tabular}{l l r r r r}
\toprule
SwVendor&Software&\multicolumn{1}{l}{\MAL}&\multicolumn{1}{l}{\STDAL}&\multicolumn{1}{l}{\EKIT}&\multicolumn{1}{l}{\textbf{Sum}}\\\midrule
Adobe & & 2 & 12 & 17 & \textbf{31}\\
&flash&0&8&10&{18}\\
&acrobat&2&4&7&{13}\\
Microsoft &  & 7 & 22 & 14 & \textbf{43}\\
&office&0&11&2&{13}\\
&int. expl.&0&4&7&{11}\\
&windows&7&6&5&{18}\\
&silverlight&0&1&0&{1}\\
Oracle & & 0&5&10&\textbf{15}\\
&java&0&5&10&{15}\\
\textbf{Sum}&&\textbf{9}&\textbf{39}&\textbf{41}&\textbf{89}\\
\bottomrule
\end{tabular}
\end{table}
reports the counts of exploited software for each product type.
\texttt{Microsoft} vulnerabilities alone make up for more than half the exploits traded as \STDAL\ products (56\%); unsurprisingly,  vulnerabilities in \texttt{Oracle} and \texttt{Adobe} products, as well as Internet Explorer vulnerabilities, are prevalent in \EKIT\ bundles, as these products are by design exposed to Internet requests~\cite{Grier-12-CCS,Kotov-2013-ESSOS}. Exploits bundled in \MAL\ are for Windows and \texttt{Adobe} Acrobat. A Fisher Exact test rejects the null hypothesis of count uniformity ($p=0.0012$), suggesting that exploited software varies by package type.

\begin{figure}
\centering
\includegraphics[width=0.8\columnwidth]{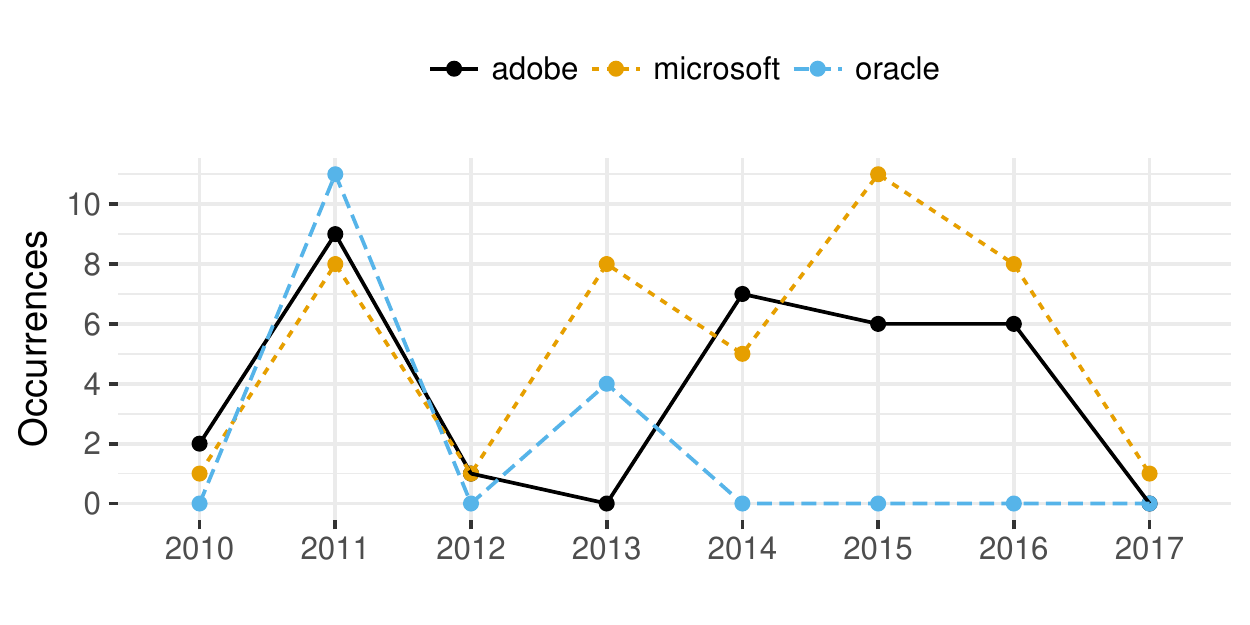}
\caption{Occurrences of exploit publication by year}
\label{fig:vendorbyyear}
\end{figure}

Figure~\ref{fig:vendorbyyear}
plots the occurrences of exploit publications by year and by software vendor.   
We observe that during the first
years of \market\ operation there is a spike in number of published exploits for all
platforms. \texttt{\texttt{Oracle}} products result as the most affected in that year, followed by \texttt{\texttt{Adobe}} and, closely, \texttt{\texttt{Microsoft}} products. This observation matches the surge around 2010-2013 of `cybercrime as a service', thoroughly reported in the scientific literature and industry in that time-frame~\cite{Grier-12-CCS,Kotov-2013-ESSOS,symantec-ekits-2011}. Interestingly,
\texttt{\texttt{Oracle}} exploits seem to plunge after 2013; this coincides with the introduction in major web browsers of plugin-blocking features,\footnote{\url{https://www.theregister.co.uk/2013/12/10/firefox\_26\_blocks\_java/}, last visit Aug 2017.} and a Java update (released in January 2013) that increases the default security settings of the plugin\footnote{\url{http://bit.ly/2r8MLz1}, last visit Aug 2017.} (e.g. triggering certificate errors as exploited by several exploit kits~\cite{Kotov-2013-ESSOS}). This  also independently supports previous findings on exploitation of Java vulnerabilities~\cite{holzinger2016depth}. Following 2013, \texttt{\texttt{Microsoft}} and \texttt{\texttt{Adobe}} exploits are publised at similar, steady rates. 
Anecdotally, we observe that the shape of the described curve resembles the \emph{Gartner Hype Cycle}\footnote{\url{http://gtnr.it/1g1Nnw0}, last visit Aug 2017.} curve, whereby after a first spike at the beginning of a new product cycle (the `\emph{Peak of Inflated Expectations}') the market experiences a relative drop (`\emph{Trough of Disillusionment}') followed by a
`plateau' where the technology reaches maturity (`\emph{Plateau of Productivity}')~\cite{Allodi-TETCS-15}. 

\subsubsection{Exploit arrival} 
\begin{figure}
\centering
\includegraphics[width=0.85\columnwidth]{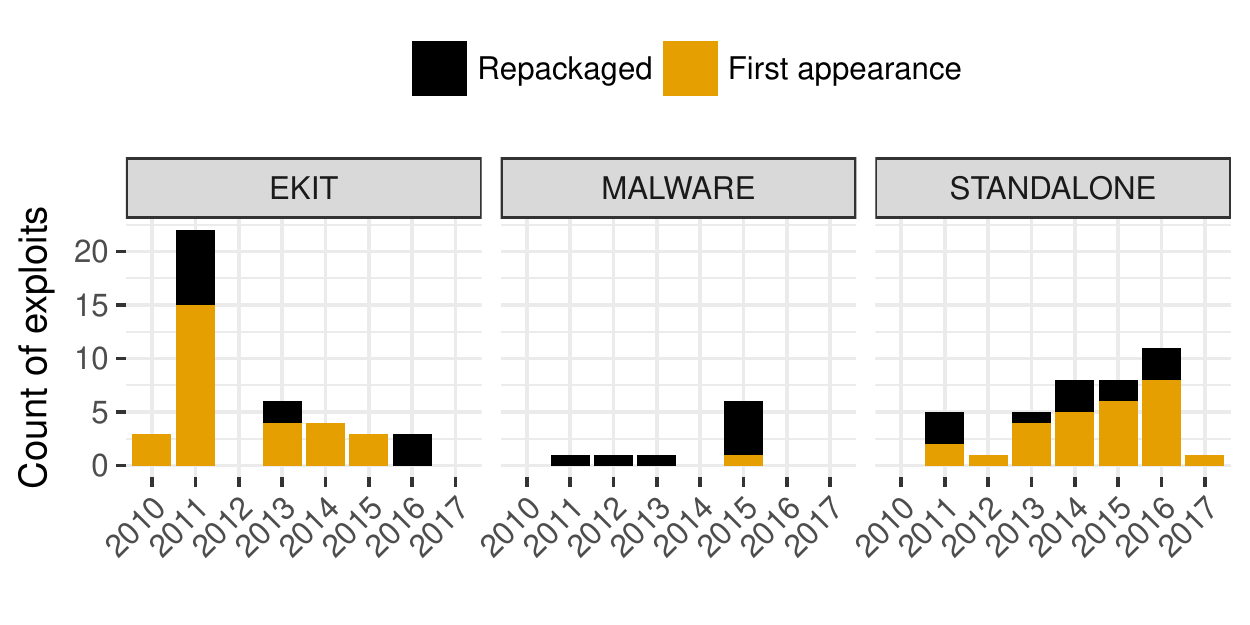}
\caption{Number of repackaged exploits by year}
\label{fig:exploitsVsYearVSFirst}
\end{figure}

Figure~\ref{fig:exploitsVsYearVSFirst}
reports the number of newly released (yellow) and repackaged (black) exploits in
each package type by year. 
The introduction of new exploits in \market\ is primarily driven by  \AL\ and \EKIT\ packages, with \MAL\ packages mainly re-introducing
already published exploits. In particular, \AL\ products seem to propose new exploits at a yearly rate of approximately 80\% for each package. \EKIT\ products introduced a significant
number of exploits in 2011 (their `debut' year on the markets~\cite{Grier-12-CCS,symantec-ekits-2011}), whereas newer exploit kits appear to embed a lower
number of exploits. This confirms previous figures whereby exploit kits are specializing to use fewer, more reliable exploits than at their original introduction~\cite{Kotov-2013-ESSOS}.
Table~\ref{tab:repack} in the Appendix
reports the evolution of repackaged exploits by \texttt{PackType}. Most exploits
first appear in \AL\ and \EKIT\ packages and re-appear
in a pack of the same type, with a few exceptions. 
Among these, \AL\
exploits seem to reappear in both \MAL\ and \EKIT\ packs, whereas \EKIT\
exploits are prevalently re-packed in other kits. \AL\ exploits seem therefore to play a role in the `innovation' process in \market; this may indicate the presence of an `exploit chain' whereby 
the most reliable and effective \AL\ exploits are selected for future inclusion in \EKIT\ products for deployment at scale.

\begin{figure}
\centering
\includegraphics[width=0.45\columnwidth]{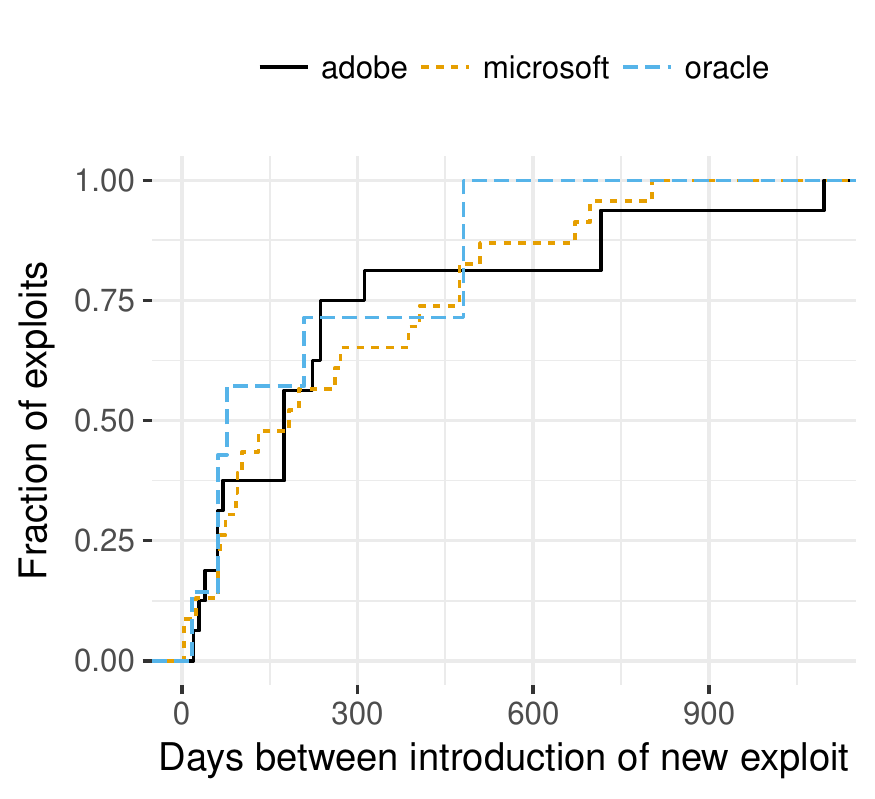}
\includegraphics[width=0.45\columnwidth]{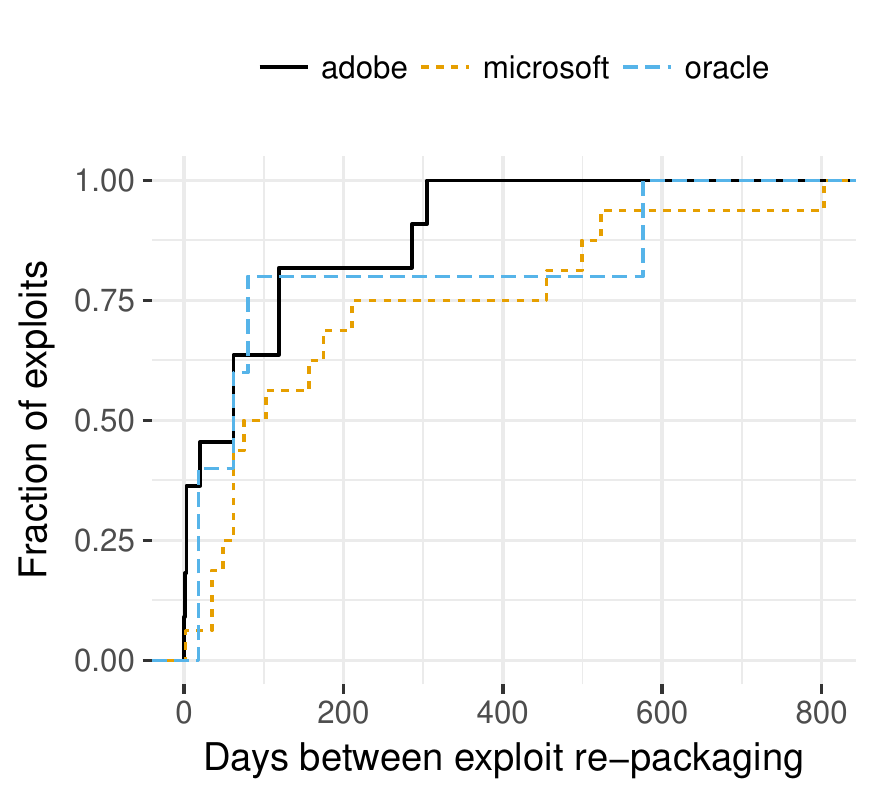}
\caption{Distribution of days between exploit introduction}
\label{fig:exploitsintro}
\end{figure}

It is interesting to evaluate the \emph{rate} at which
exploit introduction happens. A few recent studies suggest that the rate
of appearance of new exploits may be much lower than previously thought~\cite{Bilge-12-CCS,Allodi-ECIS-15}, but no account of exploit timing in the cybercrime markets currently exists.
Figure~\ref{fig:exploitsintro}
shows a breakdown by software vendor of
the distribution of days between the introduction of new exploits (left) and
re-packed exploits (right).
We exclude from the analysis six \EKIT\ vulnerabilities that have been
added to the respective packages as updates, but whose date of addition is not reported in the market. This leaves us with $83$ exploits, of which $n=55$ are introduced for the first time in the market.
 New exploits are introduced at similar rates
for all software vendors, with 50\% of exploits being introduced at approximately six months
intervals (175 days). The `fastest' 25\% is introduced two months (62 days) after the appearance of an exploit for the same software platform, whereas the `slowest' 25\% appears after more than a year (401 days). These figures are in sharp contrast with current assumptions made in the literature, whereby essentially all `severe' 
vulnerabilities are potentially
exploited at scale by attackers~\cite{Shahzad-2012-ICSE,Naaliel-ISSRE-14} (and must therefore be fixed immediately~\cite{Verizon-2015-pci}). On the contrary, these findings support recent evidence pointing in the opposite direction: most vulnerabilities
are not of `economic' interest for an attacker, as the addition of a new vulnerability may not lead to a significant
increase in targeted systems~\cite{nappa2015attack,Allodi-17-WAAM}; this results in significantly skewed distributions of risk per vulnerability (as empirically shown in~\cite{Nayak-2014-RAID,Allodi-ESSOS-15}, and analytically modeled in~\cite{Allodi-17-WAAM}). Exploit re-packaging (right plot in Fig.~\ref{fig:exploitsintro}) happens at significantly faster rates: 75\% of exploits are re-introduced within 184 days from first publication, indicating that their commercial
interest is short-lived. 

\begin{figure}
\centering
\includegraphics[width=0.75\columnwidth]{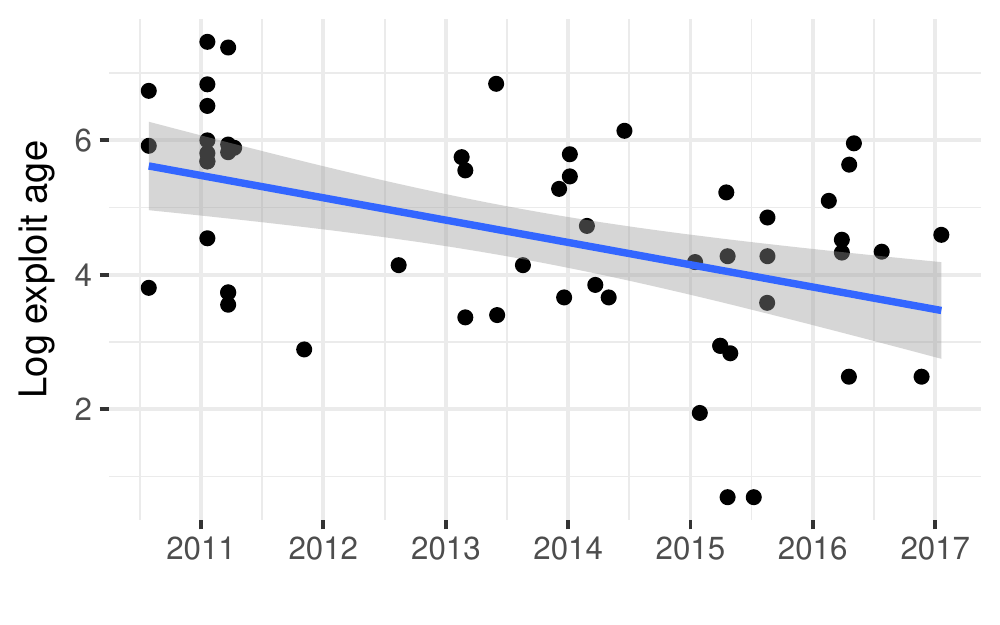}
\caption{Exploit age vs time of publication in \market}
\label{fig:ageVSyear}
\end{figure}

A different question is how `old' are exploits when they first appear on the market. We compute exploit age as the difference in days between exploit publication in the market
and publication of the corresponding CVE on the \NVD, i.e. $\texttt{ExplAge}=\texttt{PubDate}-\texttt{CVEPub}$.
Table~\ref{tab:exploitage}
\begin{table}
\small
\centering
\caption{Exploit age (days) at time of first publication}
\label{tab:exploitage}
\begin{tabular}{l r r r r rrr r}
\toprule
Type&Min&0.025p&Mean& Med.&0.975p&Max&sd & n\\\midrule
\texttt{EKIT}&1	&4&	372.48&	294	&1659.8&	1745&470.16	&25\\
\texttt{MAL}&185&185&185&185&185&185&-&1\\
\texttt{STDL}&1&8&147.34&75&549.7&934&189.66&29\\
\texttt{All}&1	&2.75	&250.36	&93&	1368.85&	1745	&359.97&	55\\
\bottomrule
\end{tabular}
\end{table}
reports the distribution of exploit age for newly introduced exploits.
The mean and median exploit age varies considerably by product type. \AL\ exploits are on average significantly younger at time of publication than other exploits ($p=0.05$ for a Wilcox rank sum test). 50\% of \AL\ exploits are published in the market
within 2.5 months (75 days) from the public disclosure date. The top 25\% (not reported in Tab.~\ref{tab:exploitage}) are published within 40 days,
and the top 2.5\% within approximately a week. The difference in exploit age for the \EKIT\ and \MAL\ categories is not statistically significant.
Whereas
some exploits do appear quickly after disclosure in \market, most exploits
take around four months from disclosure date to be published. This may
indicate that other factors such as effectiveness of older exploits~\cite{Allodi-ECIS-15},
 or delays in user system updates~\cite{nappa2015attack,sarabi2017patch}, may affect timing of appearance of a marketed exploit.
To evaluate the rate of change in time of arrival of new exploits,
Figure~\ref{fig:ageVSyear}
reports the exploit age distribution by year of publication.
We observe that the mean exploit age decreases steadily for more recent publication dates ($\beta=-0.32,\ p=0.001$), indicating that exploit vendors are becoming faster in releasing exploits for newly disclosed vulnerabilities. The coefficient of
the linear regression indicates that exploits appear at an approximately 30\% faster rate every year.

\subsubsection{Estimate of exploit prices}
Conceptually, the lower bound cost of a pack can be summarized as follows:
\begin{equation}
PackPrice_{p \in P_T} = \sum_{i=1}^n CExpl_{pi} + CDev_{p} + CDepl_{p} 
\label{eq:1}
\end{equation}
where $CExpl_{pi}$ is the cost of the $i^{th}$ exploit in package $p$, and $CDev_{p}\ \text{and}\ CDepl_{p}$ are the development and deployment (including maintenance) costs of the pack.
For example, on top of the sole exploits \EKIT s provide a web interface to control infections, as well as additional deployment services such as remote servers where the kit is hosted, or the redirection of vulnerable traffic to the kit~\cite{Kotov-2013-ESSOS,Grier-12-CCS}. Similarly, \MAL\ packages provide additional malware functionalities on top of the sole exploit. Hence, we have $CDev_{p}>0,\ CDepl_{p}>0, \forall p \in P_t$, with $t\in \{\EKIT,\ \MAL\}$. Unfortunately, 
an estimation of these costs would require an analysis of the source code of these packages~\cite{calleja2016look}, which is not publicly available.
On the other hand, \AL\ exploits are provided as-is, i.e. only the vulnerability exploit
is traded, without further embellishments or services. This sets $CDev_{p} \approx CDepl_{p} \approx 0$ for this category. This leaves us with only the
term $\sum_{i=1}^n CExpl_{pi}$ which, assuming a uniform distribution of exploit costs per package, (see discussion in Sec.~\ref{sec:analysisprocedure}), yields $CExpl_{p \in P_\texttt{STDL},i} = 1/n \cdot PackPrice_{p}$.  We therefore only report \AL\ exploit estimates.\footnote{The estimation
for all packages is reported in the Appendix, Table~\ref{tab:pricesvendorsfull}.}

Table~\ref{tab:pricesvendors}
\begin{table*}[t]
\small
\centering
\caption{(Bootstrapped) descriptive statistics of \AL\ exploit price estimates in USD}
\label{tab:pricesvendors}
\begin{minipage}{0.76\textwidth}
\footnotesize
\AL\ exploit prices are estimated on a uniform distribution by package.
To approximate the true (unknown) distribution of exploits, we perform a bootstrap of our data ($N=10000$), reported in parenthesis. 
The column $n$
reports number of exploits for that software.
The bootstrapped data does not deviate substantially
from our observations on the average. Fatter distribution tails indicate that \market\ outliers tend to bias sample statistics. Exploits are priced between 150 and 8000USD with significant differences by software.
\end{minipage}
\begin{tabular}{  p{0.13\textwidth} p{0.125\textwidth} rrr r r r r r}
\toprule
SwVendor&Software&Min & 0.025p &Mean&Median&0.975p& Max & sd&\multicolumn{1}{r}{n}\\
\midrule
\texttt{Adobe}&&75&75&879.17&1250&1500&1500&693.54&12\\
&&(75)&(100)&(1000.06)&(1040)&(1500)&(1500)&(521.91)&\\
&flash&75&75&568.75&150&1500&1500&652.3&8\\
&&(75)&(87.5)&(562.05)&(545.45)&(1300)&(1500)&(316.52)&\\
&acrobat&1500&1500&1500&1500&1500&1500&0&4\\
&&(1500)&(1500)&(1500)&(1500)&(1500)&(1500)&(0)&\\
\cmidrule{2-10}
\texttt{Microsoft}&&150&150&2801.82&2250&8000&8000&2393.09&22\\
&&(150)&(150)&(2442.13)&(2450)&(5600)&(8000)&(1601.69)&\\
&office&150&362.5&3195.45&4000&7250&8000&2504.04&11\\
&&(150)&(1605.1)&(3407.31)&(3262.5)&(5750)&(8000)&(1112.54)&\\
&int. expl.&440&459.5&3035&1850&7625&8000&3504.22&4\\
&&(440)&(440)&(3051.89)&(3000)&(8000)&(8000)&(1727.18)&\\
&windows&700&800&2366.67&2250&4687.5&5000&1458.31&6\\
&&(700)&(1100)&(2349.27)&(2327.27)&(3750)&(5000)&(658.77)&\\
&silverlight&150&150&150&150&150&150&&1\\
&&(150)&(150)&(150)&(150)&(150)&(150)&(0)&\\
\cmidrule{2-10}
\texttt{Oracle}&&25&25&1020&25&4502.5&5000&2224.89&5\\
&&(25)&(25)&(1847.02)&(1020)&(5000)&(5000)&(1981.08)&\\
&java&25&25&1020&25&4502.5&5000&2224.89&5\\
&&(25)&(25)&(1847.02)&(1020)&(5000)&(5000)&(1981.08)&\\
\bottomrule
\end{tabular}
\end{table*}
reports price estimates for exploits against different software. In parenthesis we report the bootstrapped estimation of exploit prices.
We report mean, median, standard deviation and
 95\% confidence intervals. 
Price estimates in the boostrapped sample appear to  diverge at the tails
of the distribution w.r.t the observed sample, suggesting
that outliers in the sample may bias sample statistics. 
Looking at exploits by software, we find that the most expensive 
exploits in \market\ are for \texttt{Microsoft} software,
and are priced at 2500USD on the average. Among
software from this vendor, Office and Windows exploits appear to be the most expensive with 50\% of exploits above 2000 USD, and the top 2.5\% quoted at about
 7000 and 5000 USD respectively. 
As vulnerability patching and mitigation hinder the effectiveness of an exploit in the wild~\cite{nappa2015attack}, we further evaluate whether exploit age affects exploit price estimates.
We find a negative correlation between \texttt{ExplAge} and \texttt{ExplPrice} (albeit not significant when looking only at the exploit), suggesting that exploits lose value as they age ($cor=-0.16,\ p=0.3$). We do not find evidence of dependence between exploit price and CVSS vulnerability severity.

\subsubsection{Regression analysis of exploit price estimates} To evaluate
the factors driving exploit price, we employ a set of mixed effect
linear regression models over the response variable $\log(ExplPrice)$. 
We report regression results for the following three nested models:
\begin{eqnarray}
\nonumber \text{M1:} \log(ExplPrice_i) &=&  \beta_0 + \beta_1 \log(ExplAge_i) + \epsilon_i\\
\nonumber \text{M2:} \log(ExplPrice_i) &=& \dots + \boldsymbol\beta_2 {SwVen}_i + \epsilon_i\\
\nonumber \text{M3:} \log(ExplPrice_i) &=& \dots  + \boldsymbol\beta_3 {SwVen}_i \times \log(ExplAge_i)+ \epsilon_i
 \end{eqnarray} 
Table~\ref{tab:pricereg}
\begin{table}
\small
\centering
\caption{Regression on \AL\ exploit pricing}
\label{tab:pricereg}
\begin{minipage}{0.92\columnwidth}
\footnotesize
Variables: $ExplPrice$ = price estimate of exploit; $ExplAge$ = age of exploit when advertised; $SwVen$ =  software vendor.
Exploit age
is negatively correlated with price. Depreciation rate depends on the software vendor.
\end{minipage}
\begin{tabular}{p{0.48\columnwidth} r r r }
\toprule
$\log(ExplPrice)$&\multicolumn{1}{c}{Model 1}&\multicolumn{1}{c}{Model 2}&\multicolumn{1}{c}{Model 3}\\
\midrule
c&8.080$^{^{***}}$&5.592$^{^{***}}$&10.943$^{^{***}}$\\
&(0.746)&(1.458)&(1.735)\\
$\log(ExplAge)$&-0.330$^{^{*}}$&-0.268$^{^{*}}$&-1.357$^{^{***}}$\\
&(0.129)&(0.135)&(0.234)\\
\texttt{\texttt{Adobe}}&&1.993&-4.846$^{^{*}}$\\
&&(1.395)&(2.068)\\
\texttt{\texttt{Microsoft}}&&2.662$^\dagger$&-3.483$^\dagger$\\
&&(1.375)&(1.891)\\
$\log(ExplAge) \times $ \texttt{AD}&&&1.398$^{^{***}}$\\
&&&(0.322)\\
$\log(ExplAge) \times $ \texttt{MS}&&&1.276$^{^{***}}$\\
&&&(0.272)\\
\midrule
$Var(c | ExplVen)$&2.209&1.520&1.598\\
$\text{Pseudo-}R^2$ & 0.05 & 0.28 & 0.38\\
Log-likelihood&-64.2&-60.6&-51.8\\
N&39&39&39\\
\bottomrule
\end{tabular}
\begin{minipage}{0.85\columnwidth}
\footnotesize
$(\dagger)\ p<0.1$; $(*)\ p<0.05$; $(**)\ p<0.01$; $(***)\ p<0.001$;
\end{minipage}
\end{table}
 reports the regression results. 
Coefficient estimates are consistent among models. A Variance Inflation Factors
(VIF) analysis does not indicate any significant collinearity between
the model predictors.
Log-log relationships can be interpreted as the \emph{elasticity} between dependent and independent variables. For example, in M3
the coefficient for $\log(ExplAge)$ ($-1.357$) indicates that for a 1\% increase
in the variable $ExplAge$ we can expect an average 1.4\% decrease ($1.01^{-1.357}=0.986$)
in $ExplPrice$. 
A rough quantitative approximation of the effect 
can generally be obtained by simply looking at the regression coefficients (e.g. $\beta_3=-1.357$ indicates a decrease of approximately $1.4\%$).
oefficients in M3 can only be interpreted
simultaneously with the coefficients of the interaction effect $\log(ExplAge) \times SwVen$.
We find that baseline prices for exploits vary widely by software vendor, and are
negatively correlated with the age of exploit;
\texttt{\texttt{Adobe}} and \texttt{\texttt{Microsoft}} exploits retain their value significantly longer
than \texttt{\texttt{Oracle}} exploits. This may indicate a prolonged economic interest in the
exploitation of \texttt{\texttt{Microsoft}} and \texttt{\texttt{Adobe}} vulnerabilities, a finding
consistent with related work on the persistence of vulnerabilities on end-user systems~\cite{nappa2015attack}.

Exploit vendors are a significant source of variance in price of exploit. $\text{Pseudo-}R^2$ values indicate that the models' power is adequate in explaining
the observed effect. In particular, Model 3 explains approximately 40\% 
of the variance in exploit price estimates. Importantly, all
variables of the model can be easily assessed with the sole knowledge of the vulnerability
at any point in time.






\subsection{Exploitation in the wild}
\label{sec:sym}
In this section we evaluate the effects of the identified market variables
on the exploitation in the wild of traded vulnerabilities. 
A consideration first: certain exploits may evade detection for some time, for example by means of frequent exploit repacking. On the other hand, it is unlikely for
an exploit to remain completely undercover for a long time, while in the meanwhile delivering hundreds of thousands or millions of attacks~\cite{Nayak-2014-RAID,Allodi-ESSOS-15}.
To lower uncertainty around exploit detection, we 
restrict our analysis to exploits
published in \market\ at least a year ahead of the \SYM\ data collection (i.e. before the 1st of April 2016). This coincides with the median lifetime of an exploit package in \market\ (see Fig.~\ref{fig:aliveType}), and allows \SYM\ a full calendar year to
report an exploit at scale. We consider effects before that time to be unlikely to
be caused by type II errors (i.e. no inclusion in \SYM\ despite
high attack volumes). This leaves us with $n=78$ exploits. 

\begin{figure}
\centering
\includegraphics[width=0.45\columnwidth]{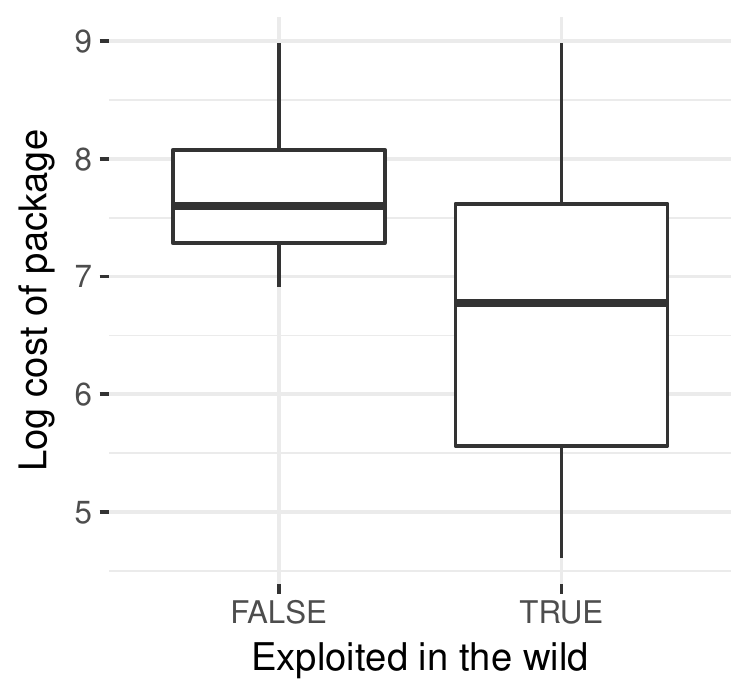}
\includegraphics[width=0.45\columnwidth]{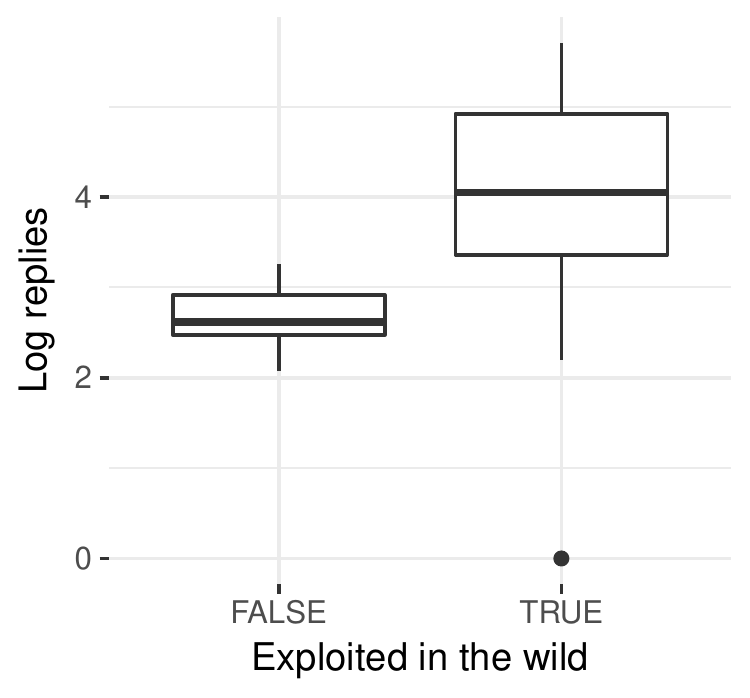}
\caption{Vulnerabilities exploited in the wild versus (left) package price; (right) replies received on the market.}
\label{fig:pricevssym}
\end{figure}

\paragraph{Package price and market activity vs exploitation}
As we are considering the effect of the \emph{acquisition \emph{and} deployment}
of an exploit by the attacker, we consider cost of package 
(as opposed to cost of exploit) because 
this reflects the upfront price the attacker needs to pay in order to deploy the attack.
Figure~\ref{fig:pricevssym}
reports the distribution of package prices (left) and replies on the market 
(right) against
exploitation in the wild. 
 Overall, we
find that exploits in \SYM\ are traded at a lower upfront price
than exploits not included in \SYM.
Similar conclusions can be drawn for the effect of market interest (as estimated by \texttt{PackActivity}) on exploitation: exploits to which
the \market\ community dedicated greater attention have a higher chance
of exploitation in the wild than exploits around which developed less market activity.
A break down by package types
does not reveal any significant interaction between the variables.
These results also support recent findings underlying the importance of the economics of the attack process in the analysis of cyber-attack scenarios~\cite{Anderson:2008,BOZORGI-etal-10-SIGKDD,eeten-2008-oecd,Nayak-2014-RAID}, and in the development
of sensible cybercrime-based measures for risk of cyber-attack~\cite{holt2016examining,Allodi-13-IWCC}. 


\paragraph{Vulnerability severity} 
Previous studies revealed the lack of correlation between technical
vulnerability characteristics and exploitation~\cite{BOZORGI-etal-10-SIGKDD}. The consideration of additional factors, such as presence of exploit in the black markets~\cite{Allodi-2014-TISSEC}, is often advised by experts and best practices to obtain more significant tests for actual exploitation~\cite{holm2015expert}. 
Following~\cite{Allodi-2014-TISSEC}, we categorize vulnerabilities
in two categories defined by the respective CVSS severity score: critical (C) ($CVSS\geq9$) and non-critical (NC) vulnerabilities ($CVSS<9$).
Table~\ref{tab:cvssSYM}
\begin{table}
\small
\centering
\caption{CVSS category vs SYM}
\label{tab:cvssSYM}
\begin{tabular}{lrrr}
\toprule
&\multicolumn{2}{c}{CVSS Category}\\
\cmidrule{2-3}
&C&NC&Sum\\\midrule
Not exploited&4&8&12\\
Exploited&53&13&66\\
Sum&57&21&78\\
\bottomrule
\end{tabular}
\end{table}
reports the corresponding distributions against \SYM. Supporting previous research findings on Exploit Kits alone~\cite{Allodi-2014-TISSEC}, 
we find that critical vulnerabilities traded in the cybercrime markets have a higher chance of exploitation in the wild (93\% in our sample) than non-critical vulnerabilities (62\%, $p=0.0021$).

\subsubsection{Regression analysis of exploitation in the wild} To more rigorously evaluate the correlation between the identified
market and vulnerability variables and exploitation in the wild, we select three logit regression models of the following form over the binary response
variable \SYM:
\begin{eqnarray}
\nonumber\text{M1: } SYM_i &=& \beta_0 +\beta_1 \log(PackActivity_i) + \epsilon_i\\
\nonumber\text{M2: } SYM_i &=& \dots + \beta_2 \log(PackPrice_i) + \epsilon_i\\
\nonumber\text{M3: } SYM_i &=& \dots +\beta_3\ \text{CVSS:NC}_i + \epsilon_i
\end{eqnarray}
where $SYM$ indicates presence or absence of exploit at scale; 
$PackActivity$ is the number of replies received by the product advert; $PackPrice$
is the upfront price to pay to obtain the exploit;
and $CVSS:NC$
is the CVSS categorization of the vulnerability severity as non-critical. 
Regression results are reported in Table~\ref{tab:regSYM}.\footnote{An OLS robustness check run on average values of regressors for each CVE (reported in the Appendix) yields equivalent results.}
\begin{table}
\small
\centering
\caption{Regression on odds of exploitation}
\label{tab:regSYM}
\begin{minipage}{0.9\columnwidth}
\footnotesize
Variables: 
$PackActivity$ = replies received in the market; $PackPrice$ = price of package; 
$CVSSCat$ = CVSS category.
Market, economic and vulnerability factors are correlated with odds of exploit at scale.
{\vspace{0.02in}}
\end{minipage}
\begin{tabular}{p{0.45\columnwidth} r r r r }
\toprule
$SYM$&\multicolumn{1}{c}{Model 1}&\multicolumn{1}{c}{Model 2}&\multicolumn{1}{c}{Model 3}\\
\midrule
c&0.245&6.056$^\dagger$&6.754$^{^{*}}$\\
&(1.526)&(3.302)&(3.086)\\
$\log(PackActivity)$&0.673$^\dagger$&0.938$^{^{*}}$&1.101$^{^{*}}$\\
&(0.383)&(0.400)&(0.451)\\
$\log(PackPrice)$&&-0.982$^{^{*}}$&-1.013$^{^{*}}$\\
&&(0.460)&(0.444)\\
\texttt{CVSS:NC}&&&-2.409$^{^{**}}$\\
&&&(0.830)\\
\midrule
$Var(c|ExplVendor)$&3.071&0.617&0.000\\
$\text{Pseudo}-R^2$ & 0.11 & 0.51 & 0.65 \\
Log-likelihood&-28.4&-25.3&-20.4\\
N&78&78&78\\
\bottomrule
\end{tabular}
\begin{minipage}{0.8\columnwidth}
\footnotesize
$(\dagger)\ p<0.1$; $(*)\ p<0.05$; $(**)\ p<0.01$; $(***)\ p<0.001$;
\end{minipage}
\end{table}
Coefficients should be interpreted as the change in the odds ratio of exploit
in the wild. For example, Model 3 indicates
that for every increase in one unit of $log(PackActivity)$ there is a three-fold increase in
odds of exploitation in the wild ($\exp(1.101)=3.01$); the
coefficient significance indicates that exploits bundled in packages around which
more market activity is developed are more likely to be detected in the wild than exploits with less activity.
Interestingly, we find that package prices also have a significant and negative impact
on odds of exploitation. This indicates that, everything else being equal, exploits bundled in more expensive packages are less likely to be detected in the wild than comparable exploits
bundled in less expensive packages.
These findings weigh favorably on the existence of a relation between market activity and exploit deployment in the wild~\cite{anderson-2012-WEIS,Allodi-2014-TISSEC,Nayak-2014-RAID}.
Vulnerability severity has also a negative impact on likelihood of exploit, indicating that
risk of exploitation for vulnerabilities in \market\ increases for critical vulnerabilities.
All models show satisfactory $\text{Pseudo-}R^2$ values, with Model 3
explaining most of the variance.

\section{Discussion}
\label{sec:discussion}


\paragraph{Exploit measures}

An important aspect of threat assessment is the consideration of exploit metrics. Current approaches often implement these by looking at the technical requirements of an attack, including the evasion of attack mitigation measures~\cite{Wang-2008-DAS} and
complexity of the attack~\cite{BOZORGI-etal-10-SIGKDD}. While a technical assessment
of the `operational' requirements of an attack can shed light on the relative ordering of attack preferences, it is hard to quantify absolute likelihoods. For example, a utility-maximizing
attacker may decide to not perform (or delay) an attack because they do not believe that there is a positive payoff, given cost of exploit
acquisition. Our analysis gives the first pointers in this direction by quantifying the relation between time, software, and exploit pricing. Importantly, this estimate only
requires readily available information on the vulnerability, and the elastic relationship
between age of exploit and price of exploit can be used to evaluate relative changes in exploit price as time passes. This directly affects current estimates of attack cost
used in risk assessment practices~\cite{Wang-2008-DAS,BOZORGI-etal-10-SIGKDD,holm2015expert}. Similarly, the \emph{update} of attackers' exploit portfolios is an important step driving
the variance in risk of attacks~\cite{Bilge-12-CCS,Allodi-ECIS-15}. 
We find that new 
exploits are introduced at rates in between two and six months, and are
approximately equal for all software. The process
driving this update remains however to uncover: 
follow up studies may look at the factors driving appearance of exploit
in the markets (e.g. by considering pre-existent exploits or software updates~\cite{nappa2015attack,Allodi-ECIS-15}). 

The dynamics of the underground markets have often 
been pointed at as an important block of overall risk of attacks, but a clear link between the two is currently missing. Whereas the problem of attack \emph{attribution} remains open (i.e. we cannot establish a causality link), this paper provides important indications on the correlation between market operations and realization of attacks. This weighs in favour of the importance of economic aspects of vulnerability
exploitation to well-informed security practices (e.g. vulnerability assessment and prioritization~\cite{holm2015expert,Verizon-2015-pci}). For example, 
our analysis of market activity and odds of exploitation in the wild reveals a significant
and positive relationship between the two. Similarly, exploits that are more expensive to acquire have lower odds
of exploitation than `cheaper' exploits. This information is often ignored in risk-assessment studies~\cite{holm2015expert}, and condensed metrics for vulnerability assessment
are used instead~\cite{Verizon-2015-pci,Naaliel-ISSRE-14}. Whereas existing vulnerability
metrics are known to not correlate to attacks in the wild~\cite{Allodi-2014-TISSEC,BOZORGI-etal-10-SIGKDD}, we find that they do once the effect of market inclusion is considered as well. Importantly, this provides a useful tool for a first evaluation of risk of exploit
without insights from the cybercrime markets other than whether the vulnerability is present~\cite{Allodi-2014-TISSEC}. A more precise estimation can then be obtained by measuring market activity around the packages embedding the exploit. These results can be factored in current best practices for vulnerability risk management and exploit mitigation~\cite{Wang-2008-DAS,MANA-WING-11-TSE,first-2015-cvss3}.

\paragraph{Vulnerability economics}

Previous studies in the literature highlighted the operations of criminal markets
for drugs, arms and pornography~\cite{soska2015measuring}, and for the 
monetization of stolen information resulting from an attack~\cite{hao2015drops}. 
However, little insight exists on the markets that, as opposed to (re)selling
the result of an attack, trade the technology that enables the attack in the first place.
A few estimates exist~\cite{Forbes-shopping-zero-days}, but are mostly 
based on 0-day price allegations, vary widely, and their relevance for the overall risk of attack remains  unclear~\cite{Bilge-12-CCS,ruohonen2016trading}. The scientific community long
discussed on the idea of building `legitimate vulnerability markets'~\cite{Ozment-2004-WEIS,Anderson:2008,kannan2005market}, and the result is the institution of a few
legitimate exploit markets~\cite{ruohonen2016trading} and of several `bug bounty' initiatives
that reward security researchers for the disclosure of new vulnerabilities~\cite{Finifter-2013-Usenix}, and discourage the participation in the underground economy. In this vein
it is interesting to observe that the prices of modern bug-bounty programs are in line or below those we identify on \market. For example, \cite[Tab.~4 pp281]{Finifter-2013-Usenix}
reports that the majority of vulnerability prizes awarded by Google in their Chrome Vulnerability Reward Program (VRP) are at or below 1000 USD, and that most
external contributors to the program (i.e. vulnerability researchers) earn between
500 and 1000 US dollars. The median price of an exploit in \market, showed in Table~\ref{tab:pricesvendors}, is at approximately 2000 dollars, a higher but not distant figure from those indicated in \cite{Finifter-2013-Usenix}. It is however unclear whether the resulting balance
weighs in favour of the legitimate or underground vulnerability markets: the dynamics balancing vulnerability finding (a notoriously demanding process~\cite{Miller-2007-WEIS}) and exploitation trade have not been fully investigated in the literature yet. For example, at the above rates it does not seem unlikely that vendors who sell their exploits multiple times to different buyers may still be better off participating in the cybercrime economy than moving to the 
`legitimate' vulnerability markets, as in the latter vulnerabilities can realistically  be traded only once (as the trade creates an association between the `0-day' vulnerability and identity of whom discovered it). 
 The results in this paper represent a first building block in the evaluation and enhancement of current legitimate vulnerability markets, to
foster the responsible disclosure of vulnerabilities and attract skillful researchers
away from criminal markets.

Our exploit price estimates provide additional insights on the effect
of different criminal business models on exploit pricing and, therefore, accessibility of attack. With reference to 
Tab.~\ref{tab:pricesvendorsfull} in the Appendix, and despite the upwards bias of the estimate (Eq.~\ref{eq:1}), \EKIT\ exploits are priced, per unit, significantly below exploits in other package types. This effect is driven by the higher number of exploits bundled in \EKIT\ (ref. Table~\ref{tab:packprices}), and underlines how 
the different business model employed by \EKIT\ services may allow exploit vendors to drastically reduce exploit development and deployment costs. Lower prices may make these tools more accessible
to `wanna-be-criminals', and therefore generate more attacks overall. 
This suggests that the criminal business model may play a central role in the diffusion of cyber-attacks, and calls for additional studies characterizing this effect.
Further, we find that the studied market shows clear signs of expansion, with a growing number of vendors, exploits, attack products, and generally inflating package prices. This indicates that market activity is unlikely to stop in the near future, and that attacker economics will likely play an increasingly more relevant role in the cybersecurity scenario.

\section{Conclusions}
\label{sec:conclusions}
In this paper we presented the first quantitative account of exploit pricing and market effects on exploitation in the wild. Our findings quantify a strong correlation between market activities and likelihood of exploit. We find that the analyzed market shows signs of expansion, and that exploit-as-a-service models
may allow for drastic cuts in exploit development costs. Further, we find that exploit prices are aligned with or above those of `legitimate' vulnerability markets, supporting work on the identification of incentives for responsible
vulnerability disclosure and attack economics.

\section{Acknowledgments}
This line of work started in collaboration with Prof. Fabio Massacci at the University of Trento, Italy, and Prof. Julian Williams at Durham Business School, UK, whom the author thanks for their invaluable insights and feedback. This work has been supported by the 
\grantsponsor{NWO}{NWO}{https://www.nwo.nl} through the SpySpot project no. \grantnum{NWO}{628.001.004}.

\bibliographystyle{ACM-Reference-Format}
 \balance
\bibliography{short-names.bib,security-common.bib,bib.bib} 
\cleardoublepage
\newpage
\section*{Appendix}

\section*{Extended description of data fields}

In the following we report a detailed list of the collected data fields.

\textbf{\texttt{CVE}}. The Common Vulnerability and Exposures ID assigned by the MITRE corporation to disclosed software vulnerabilities. 

\textbf{\texttt{CVEPub}}. The date of vulnerability  publication on \NVD.

\textbf{\texttt{ExplVen}}. The anonymized market identifier of the user who posts
the advert on the market.

\textbf{\texttt{ExplVenReg}}. The date of vendor
registration on the forum.

\textbf{\texttt{Pack}}. The exploit package advertised by the seller on the market. Packages bundle one or more vulnerability exploits.

\textbf{\texttt{PubDate}}. The date of package publication in \market.

\textbf{\texttt{PackType}}. The type of exploit package. The categorization emerges from the \texttt{ExplVen} package description. 




\textbf{\texttt{PackPrice}}. The upfront price that the customer has to pay to acquire or rent the package. For rental packages we compute the mean rental price for a rental period of 3 weeks, in line
with previous work on the duration of malware delivery campaigns and user infections~\cite{huang2014socio}. All prices in \market\ are reported in USD.

\textbf{\texttt{PackActivity}}. The \market\ activity around an exploit package expressed 
as the number of replies received by the product advert on the market.

\textbf{\texttt{PackDeath}}. The date when \market\ activity around the package stops.

\textbf{\texttt{ExplPrice}}. Estimation of cost of a single exploit. When only one exploit is bundled in a package, this corresponds to the package price, otherwise we provide an estimation. Further considerations are given in Sec.~\ref{sec:analysisprocedure}.





\textbf{\texttt{SwVen}}. The vendor of the vulnerable software affected by the published exploit.

\textbf{\texttt{Sw}}. The vulnerable software or platform.


\textbf{\texttt{CVSS}}. The severity of the exploited vulnerability as expressed by the Common Vulnerability
Scoring System, the standard framework for evaluation of software vulnerability characteristics~\cite{first-2015-cvss3}.

\textbf{\SYM}. Presence or absence of an exploit in the wild at scale.

\section*{SYM regression}

Table~\ref{tab:symols} below reports standard OLS coefficients using
the average value of the fixed effects for each CVE as regressors. Results are equivalent
to those shown by the mixed model in Table~\ref{tab:regSYM}. 
\begin{table}[h!]
\footnotesize
\caption{OLS regression of exploitation in the wild}
\label{tab:symols}
\begin{tabular}{p{0.45\columnwidth} r r r r }
\toprule
\emph{SYM}&\multicolumn{1}{c}{Model 1}&\multicolumn{1}{c}{Model 2}&\multicolumn{1}{c}{Model 3}\\
\midrule
c&-2.062&5.652&6.846$^\dagger$\\
&(1.479)&(3.580)&(3.959)\\
$\log(\texttt{PackActivity})$&1.108$^{^{*}}$&1.436$^{^{**}}$&1.593$^{^{*}}$\\
&(0.460)&(0.549)&(0.627)\\
$\log(PackPrice)$&&-1.184$^{^{*}}$&-1.299$^{^{*}}$\\
&&(0.546)&(0.589)\\
\texttt{CVSS:NC}&&&-2.184$^\dagger$\\
&&&(1.124)\\
\midrule
McFadden R-sq.&0.2&0.4&0.5\\
Log-likelihood&-17.7&-14.1&-11.9\\
N&51&51&51\\
\bottomrule
\end{tabular}
\end{table}


\begin{table}[h]
\small
\centering
\caption{CVSS Access Complexity by package}
\label{tab:cvsspack}
\begin{tabular}{l r r r r}
\toprule
&H&L&M&Sum\\
\midrule
ADB&0&1&1&2\\
ADB2&0&0&1&1\\
ADB3&0&1&1&2\\
BANN&0&1&1&2\\
BOT&0&1&0&1\\
DOC&0&0&1&1\\
DOCPPT&0&0&1&1\\
DROP&0&2&0&2\\
DROP2&0&2&1&3\\
ELEN&1&2&8&11\\
FFLA&0&0&2&2\\
FLASH&0&1&2&3\\
GLUE&0&0&1&1\\
GLUE2&0&0&1&1\\
IE11&0&0&1&1\\
IE311&0&0&1&1\\
IEG8&1&0&0&1\\
JAVA&0&3&1&4\\
JAVA2&0&1&0&1\\
KATR&1&5&5&11\\
LPE&0&1&0&1\\
MSW&0&0&1&1\\
MSW2&0&0&2&2\\
NEUT&0&2&0&2\\
OFF&0&0&1&1\\
PAC&0&0&1&1\\
PDF&0&1&1&2\\
PRIV&1&2&1&4\\
PRIV1&0&1&0&1\\
R0&0&2&0&2\\
RIG&0&5&5&10\\
SILV&0&0&1&1\\
VIS&0&1&0&1\\
WMI4&0&0&4&4\\
XP&0&0&1&1\\
ZOMB&0&0&3&3\\
Sum&4&35&50&89\\
\bottomrule
\end{tabular}
\end{table}

\begin{table*}
\caption{CWE counts by package}
\label{tab:cwebypackage}
\begin{minipage}{0.35\textwidth}
\footnotesize
CWEs are uniformly distributed among packages, i.e. each
pack contains vulnerabilities of the same or comparable type. This
indicates that within a single package, exploit development cost 
are comparable and significantly skewed distributions of costs are
unlikely.\\
1. Buf. Err. (CWE-119)\\
2. Code Inj. (CWE-94)\\
3. Data Handl. (CWE-19)\\
4. Input Val. (CWE-20)\\
5. Ins. Inf, (NVD-CWE-noinfo)\\
6. Link Following (CWE-59)\\
7. NONE\\
8. Num. Err. (CWE-189)\\
9. OS Comm. Inj. (CWE-78)\\
10. (NVD-CWE-Other)\\
11. Path Trav. (CWE-22)\\
12. Perm., Priv., Acc. Cntrl (CWE-264)\\
13. Res. Mngmt Err (CWE-399)\\
14. Use Af. Free (CWE-416)
\end{minipage}
\footnotesize
\begin{tabular}{lrrrrrrrrrrrrrrr}
\toprule
&1&2&3&4&5&6&7&8&9&10&11&12&13&14&Sum\\
\midrule
ADB&0&0&0&0&1&0&0&1&0&0&0&0&0&0&2\\
ADB2&0&0&0&0&0&0&0&1&0&0&0&0&0&0&1\\
ADB3&0&0&0&0&1&0&0&1&0&0&0&0&0&0&2\\
BANN&0&0&0&0&0&0&0&1&0&1&0&0&0&0&2\\
BOT&0&0&0&1&0&0&0&0&0&0&0&0&0&0&1\\
DOC&0&0&0&1&0&0&0&0&0&0&0&0&0&0&1\\
DOCPPT&0&0&0&0&0&0&0&0&0&0&0&0&1&0&1\\
DROP&0&0&0&0&0&0&0&0&0&0&0&2&0&0&2\\
DROP2&1&0&0&1&0&0&0&0&0&0&0&1&0&0&3\\
ELEN&2&2&0&0&3&0&1&1&1&0&0&0&1&0&11\\
FFLA&0&0&0&0&2&0&0&0&0&0&0&0&0&0&2\\
FLASH&2&0&0&0&0&0&0&1&0&0&0&0&0&0&3\\
GLUE&0&1&0&0&0&0&0&0&0&0&0&0&0&0&1\\
GLUE2&0&1&0&0&0&0&0&0&0&0&0&0&0&0&1\\
IE11&1&0&0&0&0&0&0&0&0&0&0&0&0&0&1\\
IE311&0&1&0&0&0&0&0&0&0&0&0&0&0&0&1\\
IEG8&1&0&0&0&0&0&0&0&0&0&0&0&0&0&1\\
JAVA&0&0&0&0&3&0&0&0&1&0&0&0&0&0&4\\
JAVA2&0&0&0&0&1&0&0&0&0&0&0&0&0&0&1\\
KATR&1&2&0&1&5&0&0&0&1&0&0&0&1&0&11\\
LPE&0&0&0&0&0&0&0&0&0&0&0&1&0&0&1\\
MSW&1&0&0&0&0&0&0&0&0&0&0&0&0&0&1\\
MSW2&0&0&1&0&0&0&0&0&0&1&0&0&0&0&2\\
NEUT&0&0&0&0&2&0&0&0&0&0&0&0&0&0&2\\
OFF&0&0&0&1&0&0&0&0&0&0&0&0&0&0&1\\
PAC&0&1&0&0&0&0&0&0&0&0&0&0&0&0&1\\
PDF&0&0&0&0&1&0&0&1&0&0&0&0&0&0&2\\
PRIV&0&1&0&0&3&0&0&0&0&0&0&0&0&0&4\\
PRIV1&0&0&0&1&0&0&0&0&0&0&0&0&0&0&1\\
R0&0&0&0&0&0&0&0&0&0&0&0&2&0&0&2\\
RIG&2&1&0&0&4&0&0&1&0&1&0&0&0&1&10\\
SILV&0&0&0&1&0&0&0&0&0&0&0&0&0&0&1\\
VIS&0&0&0&0&0&0&0&0&0&0&0&1&0&0&1\\
WMI4&2&2&0&0&0&0&0&0&0&0&0&0&0&0&4\\
XP&1&0&0&0&0&0&0&0&0&0&0&0&0&0&1\\
ZOMB&0&1&0&0&0&0&0&1&1&0&0&0&0&0&3\\
Sum&14&13&1&7&26&0&1&9&4&3&0&7&3&1&89\\
\bottomrule
\end{tabular}
\end{table*}

\begin{table*}[h]
\centering
\footnotesize
\caption{History of re-packaged vulnerabilities}
\label{tab:repack}
\begin{tabular}{l r l l l l l l}
\toprule
CVE&no.&Sw&SwVendor&Type 1&Type 2&Type 3&Type 4\\\midrule
2015-8651&4&flash&adobe&\AL&\AL&\AL&\EKIT\\
2010-0188&4&acrobat&adobe&\EKIT&\EKIT&\MAL&\MAL\\
2012-1864&3&windows&microsoft&\AL&\MAL&\MAL&\\
2015-1701&3&windows&microsoft&\AL&\MAL&\MAL&\\
2010-4452&3&java&oracle&\EKIT&\EKIT&\AL&\\
2006-0003&3&int.\_expl.&microsoft&\EKIT&\EKIT&\EKIT&\\
2010-1885&3&windows&microsoft&\EKIT&\EKIT&\EKIT&\\
2016-1019&2&flash&adobe&\AL&\EKIT&&\\
2013-2729&2&acrobat&adobe&\AL&\AL&&\\
2013-0640&2&acrobat&adobe&\AL&\AL&&\\
2015-2545&2&office&microsoft&\AL&\AL&&\\
2015-0057&2&windows&microsoft&\AL&\MAL&&\\
2013-3660&2&windows&microsoft&\AL&\MAL&&\\
2011-0611&2&acrobat&adobe&\EKIT&\EKIT&&\\
2010-0886&2&java&oracle&\EKIT&\AL&&\\
2008-2463&2&office&microsoft&\EKIT&\EKIT&&\\
2015-0336&2&flash&adobe&\EKIT&\AL&&\\
2013-3918&2&int.\_expl.&microsoft&\AL&\AL&&\\
2015-2419&2&int.\_expl.&microsoft&\AL&\EKIT&&\\
2014-6332&2&windows&microsoft&\EKIT&\AL&&\\
2010-0840&2&java&oracle&\EKIT&\AL&&\\
2011-3544&2&java&oracle&\AL&\EKIT&&\\
2012-0158&2&office&microsoft&\AL&\AL&&\\
\bottomrule
\end{tabular}
\end{table*}

\begin{table*}[t]
\centering
\footnotesize
\caption{(Bootstrapped) descriptive statistics of exploit price estimates in USD by software}
\label{tab:pricesvendorsfull}
\begin{minipage}{0.85\textwidth}
\footnotesize
We compute expected exploit prices by considering a uniform distribution
of exploit costs by package.
To approximate the true (unknown) distribution of exploits per package
and software, we perform a bootstrap of our data ($N=10000$). Estimates for \MAL\ and \EKIT\
exploits are only indicative as development costs of the package can not be accounted for.
We report descriptive
statistics of the original and of the bootstrapped sample means in parenthesis.
The column $n$
reports number of exploits for that software in the respective package type.
The bootstrapped data does not deviate substantially
from our observations on the average. Fatter distribution tails indicate that \market\ outliers tend to bias sample statistics. 
\texttt{Microsoft} exploits are on average the most valuable in the market irrespective of package. MS Windows
and Office exploits are consistently the most expensive. \texttt{Adobe} and \texttt{Oracle} are closely second and third. \EKIT\ prices by exploit are substantially
lower than for \AL\ and \MAL\ exploits.
\end{minipage}
\begin{tabular}{ l l l rrr r r r r r}
\toprule
\multicolumn{1}{l}{Pack Type}&SwVendor&Software&Min & 0.025p &Mean&Median&0.975p& Max & sd&\multicolumn{1}{r}{n}\\
\midrule
\multirow{18}{*}{\EKIT}
&\texttt{Adobe}&&13.64&13.64&98.24&31.67&333.33&333.33&111.28&17\\
&&&(13.64)&(13.64)&(135.79)&(106.86)&(333.33)&(333.33)&(90.38)&\\
&&flash&31.67&31.67&91.86&31.67&299.24&333.33&105.33&10\\
&&&(31.67)&(31.67)&(132.99)&(98.7)&(333.33)&(333.33)&(89.9)&\\
&&acrobat&13.64&13.64&107.36&13.64&310.61&333.33&127.34&7\\
&&&(13.64)&(13.64)&(138.59)&(114.55)&(333.33)&(333.33)&(90.77)&\\
&\texttt{Microsoft}&&13.64&13.64&115.71&120.83&284.09&333.33&96.76&14\\
&&&(13.64)&(13.64)&(113.81)&(111.44)&(232.32)&(333.33)&(62.23)&\\
&&office&13.64&17.84&97.73&97.73&177.61&181.82&118.92&2\\
&&&(13.64)&(13.64)&(98.44)&(97.73)&(181.82)&(181.82)&(63.88)&\\
&&int. expl.&13.64&16.34&97.47&120.83&181.82&181.82&71.89&7\\
&&&(13.64)&(13.64)&(94.92)&(97.47)&(181.82)&(181.82)&(38.63)&\\
&&windows&13.64&15.44&148.45&181.82&318.18&333.33&130.6&5\\
&&&(13.64)&(13.64)&(146.27)&(148.45)&(333.33)&(333.33)&(66.71)&\\
&\texttt{Oracle}&&13.64&13.64&87.32&106.67&181.82&181.82&68.78&10\\
&&&(13.64)&(13.64)&(94.75)&(92.26)&(181.82)&(181.82)&(39.1)&\\
&&java&13.64&13.64&87.32&106.67&181.82&181.82&68.78&10\\
&&&(13.64)&(13.64)&(94.75)&(92.26)&(181.82)&(181.82)&(39.1)&\\
\midrule
\multirow{8}{*}{\MAL}
&\texttt{Adobe}&&420&421.75&455&455&488.25&490&49.5&2\\
&&&(420)&(420)&(455.35)&(455)&(490)&(490)&(26.71)&\\
&&acrobat&420&421.75&455&455&488.25&490&49.5&2\\
&&&(420)&(420)&(455.35)&(455)&(490)&(490)&(26.71)&\\
&\texttt{Microsoft}&&333.33&333.33&1357.14&1500&3625&4000&1303.23&7\\
&&&(333.33)&(333.33)&(1518.3)&(1375)&(4000)&(4000)&(745.71)&\\
&&windows&333.33&333.33&1357.14&1500&3625&4000&1303.23&7\\
&&&(333.33)&(333.33)&(1518.3)&(1375)&(4000)&(4000)&(745.71)&\\
\midrule
\multirow{18}{*}{\AL}
&\texttt{Adobe}&&75&75&879.17&1250&1500&1500&693.54&12\\
&&&(75)&(100)&(1000.06)&(1040)&(1500)&(1500)&(521.91)&\\
&&flash&75&75&568.75&150&1500&1500&652.3&8\\
&&&(75)&(87.5)&(562.05)&(545.45)&(1300)&(1500)&(316.52)&\\
&&acrobat&1500&1500&1500&1500&1500&1500&0&4\\
&&&(1500)&(1500)&(1500)&(1500)&(1500)&(1500)&(0)&\\
&\texttt{Microsoft}&&150&150&2801.82&2250&8000&8000&2393.09&22\\
&&&(150)&(150)&(2442.13)&(2450)&(5600)&(8000)&(1601.69)&\\
&&office&150&362.5&3195.45&4000&7250&8000&2504.04&11\\
&&&(150)&(1605.1)&(3407.31)&(3262.5)&(5750)&(8000)&(1112.54)&\\
&&int. expl.&440&459.5&3035&1850&7625&8000&3504.22&4\\
&&&(440)&(440)&(3051.89)&(3000)&(8000)&(8000)&(1727.18)&\\
&&windows&700&800&2366.67&2250&4687.5&5000&1458.31&6\\
&&&(700)&(1100)&(2349.27)&(2327.27)&(3750)&(5000)&(658.77)&\\
&&silverlight&150&150&150&150&150&150&&1\\
&&&(150)&(150)&(150)&(150)&(150)&(150)&(0)&\\
&\texttt{Oracle}&&25&25&1020&25&4502.5&5000&2224.89&5\\
&&&(25)&(25)&(1847.02)&(1020)&(5000)&(5000)&(1981.08)&\\
&&java&25&25&1020&25&4502.5&5000&2224.89&5\\
&&&(25)&(25)&(1847.02)&(1020)&(5000)&(5000)&(1981.08)&\\
\bottomrule
\end{tabular}
\end{table*}



\end{document}